\documentclass[12pt]{article}
\pdfoutput=1
\usepackage[english]{babel}
\usepackage{fancyhdr}
\usepackage{amsmath}
\usepackage{amssymb}
\usepackage{amsfonts}
\usepackage{psfrag}
\usepackage[applemac]{inputenc}
\usepackage{graphicx,wrapfig}
\usepackage[bf,footnotesize]{caption2}
\usepackage{pstricks}

\usepackage{cite}

\def\m@th{\mathsurround=0pt }
\def\leftrightarrowfill{$\m@th \mathord\leftarrow \mkern-6mu
	\cleaders\hbox{$\mkern-2mu \mathord- \mkern-2mu$}\hfill
	\mkern-6mu \mathord\rightarrow$}

\def\overleftrightarrow#1{\vbox{\ialign{##\crcr
	\leftrightarrowfill\crcr\noalign{\kern-1pt\nointerlineskip}
	$\hfil\displaystyle{#1}\hfil$\crcr}}}

\newcommand{\be}{\begin{equation}}
\newcommand{\ee}{\end{equation}}

\newcommand{\Tr}{\mathop{\rm Tr}}
\def\I{\rm 1\kern-.24em l}  

\newcommand{\newc}{\newcommand}

\newc{\gsim}{\lower.7ex\hbox{$\;\stackrel{\textstyle>}{\sim}\;$}}
\newc{\lsim}{\lower.7ex\hbox{$\;\stackrel{\textstyle<}{\sim}\;$}}
\newc{\ie}{{\it i.e.}}
\newc{\etal}{{\it et al.}}
\newc{\mev}{\hbox{\rm\,MeV}}
\newc{\gev}{\hbox{\rm\,GeV}}
\newc{\tev}{\hbox{\rm\,TeV}}
\newc{\xpb}{\hbox{\rm\, pb}}
\newc{\xfb}{\hbox{\rm\, fb}}

\newc{\mtop}{m_t}
\newc{\mbot}{m_b}
\newc{\mz}{M_Z}
\newc{\mw}{M_W}
\newc{\alphasmz}{\alpha_s(M_Z)}
\newc{\swsq}{\sin^2\theta_W}
\newc{\cwsq}{\cos^2\theta_W}
\newc{\tw}{\tan\theta_W}
\newc{\cw}{\cos\theta_W}
\newc{\sw}{\sin\theta_W}
\newc{\BR}{\hbox{\rm BR}}
\newc{\zbb}{Z\to b\bar}
\newc{\Gb}{\Gamma (Z\to b\bar b)}
\newc{\Gh}{\Gamma (Z\to \hbox{\rm hadrons})}
\newc{\sgn}{\mbox{sgn}}

\newcounter{mysubequation}[equation]

\newcommand{\GeV}{\,\mathrm{GeV}}

\newcommand{\mc}{M_{q^*}}
\newcommand{\gslash}[1]{\slash{ \hspace{-2.8mm} #1}}

\def\beq{\begin{equation}}
\def\eeq{\end{equation}}
\def\bea{\begin{eqnarray}}
\def\eea{\end{eqnarray}}

\def\slashchar#1{\setbox0=\hbox{$#1$}           
\dimen0=\wd0                                 
\setbox1=\hbox{/} \dimen1=\wd1               
\ifdim\dimen0>\dimen1                        
   \rlap{\hbox to \dimen0{\hfil/\hfil}}      
   #1                                        
\else                                        
   \rlap{\hbox to \dimen1{\hfil$#1$\hfil}}   
   /                                         
\fi}                                         %
\catcode`@=11
\long\def\@caption#1[#2]#3{\par\addcontentsline{\csname
ext@#1\endcsname}{#1}{\protect\numberline{\csname
the#1\endcsname}{\ignorespaces #2}}\begingroup
 \small
 \@parboxrestore
 \@makecaption{\csname fnum@#1\endcsname}{\ignorespaces #3}\par
\endgroup}
\catcode`@=12

\begin{document}

\baselineskip=18pt

\setcounter{footnote}{0}
\setcounter{figure}{0}
\setcounter{table}{0}

\begin{titlepage}
\begin{flushright}
UAB-FT--647\\
June 2008
\end{flushright}
\vspace{.3in}

\begin{center}
\vspace{1cm}

{\Large \bf  
Top Quark Compositeness: Feasibility and Implications}

\vspace{.8cm}

{\bf Alex  Pomarol and Javi Serra}

\vspace{.5cm}

\centerline{{\it  IFAE, Universitat Aut\`onoma de Barcelona, 08193 Bellaterra, Barcelona}}

\end{center}
\vspace{.8cm}

\begin{abstract}
\medskip
\noindent
In models of electroweak symmetry breaking in which the SM fermions get their masses by mixing with composite states, it is natural to expect the top quark to show properties of compositeness. We study the phenomenological viability of having a mostly composite top. The strongest constraints are shown to mainly come from one-loop contributions to the $T$-parameter. Nevertheless, the presence of light custodial partners weakens these bounds, 
allowing in certain cases for a high degree of top compositeness. We find regions in the parameter space in which the $T$-parameter receives moderate positive contributions, favoring the electroweak fit of this type of models. We also study the implications of having a composite top at the LHC, focusing on the process $pp\rightarrow t\bar tt\bar t(b\bar b)$ whose cross-section is enhanced at high-energies.

\end{abstract}

\bigskip
\bigskip

\end{titlepage}


\section{Introduction}
\label{intro}
Unraveling  the  origin of the electroweak symmetry breaking (EWSB) is the main priority
of the LHC. One possibility,  inspired by  QCD, is that  EWSB occurs in a  
new strong sector at energies of few TeV.
Examples of this realization are Technicolor  models \cite{TC}
and  composite Higgs scenarios \cite{GK}.
More recently, due to the connection between strongly-coupled theories
and gravity on warped extra dimensions, 
these scenarios have been studied in the framework of five-dimensional theories
(see for example, Refs.~\cite{Csaki:2003zu, Agashe:2004rs}).

In all these examples  the SM fields that get masses from EWSB
must at least be coupled to this new (strong) sector 
with a strength  proportional to their masses.
This suggests that the 
top quark is  the SM field  with the largest coupling to the new sector, and  therefore  the most  sensitive to new physics. If this is the case,
the top is the most likely  SM fermion to show signals  
of compositeness.
Knowing the degree of compositeness of the top 
is then  very   important  to understand 
the physics lying  beyond the SM.

The aim of this paper is twofold.
First, we want to   study the  viability of having a    top quark being mostly a composite state.
We will study this possibility  in a framework, inspired by extra-dimensional models,  in which   the SM fermions are a mixture of 
elementary and composite states, with a 
mixing angle proportional to  $\sqrt{m_f}$, where $m_f$ is the fermion mass.
We will take the limit in which one of the two   chiral components  of the top
is mostly  a composite state, 
and study the phenomenological viability of this limit.
The main constraints  from present experiments will arise
from the $T$-parameter. 
We will calculate the one-loop contributions to $T$ and show
under which conditions  a  composite  top is allowed.
An important role  will be  played by   the custodial partners of the top,
the custodians,
that become light in the composite limit and    reduce significantly
the total  contribution to $T$. 
Our results will also be useful to determine  how    
 a positive contribution to $T$ can arise,  as required, in this class of models, to accommodate a large and positive  $S$-parameter. 

Secondly, we will show  how future experiments can  test the  properties of the top
and tell us about the degree of its compositeness.
We will do this by following a   model-independent 
approach, similar  to Ref.~\cite{Giudice:2007fh},
in which the top compositeness is characterized  by
few higher-dimensional operators.
We will concentrate on the study of the process $pp\rightarrow t\bar t t \bar t (b\bar b)$ 
that, for a composite top, is enhanced  at high-energies.
We will calculate the cross-section of this process  and  
show how different observables  can be used to distinguish
between   a composite and   elementary top.

The organization of the paper is as follows.
In section 2 we  present a   framework for a   composite top.
Its low-energy effective lagrangian is given in  section 3.
The experimental  constraints are presented in section 4;
we study the effects on  $Zb\bar{b}$
and the one-loop contributions to the  $T$-parameter.
We  present the regions of the parameter space in which 
a composite top is allowed.
In section 5 we  show how to study  the top properties at future experiments and 
present the calculation for  $pp\rightarrow t\bar t t \bar t (b\bar b)$.
We conclude in section 6.

\section{Framework}
\label{framework}
The framework  we want to consider is the following.
We will assume that beyond the SM there is a new sector (the BSM sector),  characterized by two parameters,  
a generic coupling $g_\rho$
and a  mass scale  $M_\rho$.
We will be mostly interested in the limit 
$1<g_\rho \lesssim  4 \pi$ such that the BSM sector  consists of 
resonances whose coupling, although large, allows us  for a perturbative expansion.
Our analysis, however,  will  be able to be extended 
to  the region  $g_\rho \sim 4 \pi$ corresponding  to a maximally strongly-coupled  BSM.
The scale $M_\rho$, in analogy with QCD, will correspond to the mass of the lightest resonance.
Examples of this class of models are strongly-coupled gauge theories in the large-$N$ limit  or extra dimensional models \cite{Csaki:2003zu, Agashe:2004rs}.

We will also assume that this new sector   is  responsible for the EWSB. 
This means that
the Goldstone bosons  $G^a$ (to be eaten by the  $W$ and $Z$) 
will    arise from the  BSM sector. 
They  can  be  parametrized by a  matrix  $\Sigma$ whose vacuum expectation value (VEV) 
breaks the EW symmetry:  
\begin{equation}
\Sigma=v\, e^{i\sigma^a G^a/v}\ ,\ \ {\rm where}\ \   v\simeq 246\ {\rm GeV}
\, .
\end{equation}
In Higgsless theories $v$ is equal to the decay constants of the Goldstones $f$ 
which   can be written as   
\begin{equation}
f=\frac{M_\rho}{g_\rho}\, .
\label{fdef}
\end{equation} 
In theories in which  the   Higgs  arises from the BSM sector
as a Pseudo-Goldstone Boson (PGB) the scale $f$, satisfying Eq.~(\ref{fdef}), 
is associated to the PGB-Higgs decay constant. 
The EW scale  $v$ is determined in these models by minimizing the Higgs potential
and one generically obtains   $v\lesssim f$ \cite{GK,Agashe:2004rs}.
To incorporate both scenarios, Higgsless and composite Higgs,
we will parametrize the deviation of $v$ from $f$  by
the dimensionless parameter $\xi$ defined by \cite{Giudice:2007fh}
\begin{equation}
\xi=\frac{v^2}{f^2}\leq 1\, . 
\end{equation} 
Electroweak precision tests (EWPT) put tight constraints on models of this  class, 
since the  BSM resonances 
induce sizable tree-level modifications of  the SM gauge propagators.
The main effects can be parametrized by 
two quantities, the $S$ and $T$  parameters \cite{Peskin:1991sw}.
The tree-level contribution to $ T$ can   vanish if  
the BSM sector is invariant under a global SU(2)$_V$  symmetry, the so-called custodial symmetry.  
For this reason, we will assume  that the BSM sector is  invariant
under a  global   SU(2)$_L\times$SU(2)$_R$
under which  the Goldstone  multiplet $\Sigma$   transforms as a ${\bf (2,2)}$.
The VEV of  $\Sigma$  will break 
SU(2)$_L\times$SU(2)$_R$ down to the diagonal subgroup corresponding to the custodial
symmetry.
We will further impose that the BSM sector is also invariant under the  discrete  symmetry $P_{LR}$ that interchanges    $L\leftrightarrow R$.
As we will see later, this extra parity is crucial to avoid large corrections to $Zb\bar b$ \cite{Agashe:2006at}.
Under these assumptions the only important tree-level  constraint  on this class of models comes from the $S$-parameter.  
In   extra dimensional models in which $S$ is calculable 
one  finds, barring cancellations,   the bound
$M_\rho\gtrsim 2.3$ TeV
\footnote{Similar bound is obtained if we  use the QCD experimental data  to extract
the value of $S$ \cite{Peskin:1991sw}.}
\cite{Agashe:2004rs}, or equivalently,
\begin{equation}
f \gtrsim  500\ {\rm GeV}\ \ \   (\xi\lesssim 1/4)\  \ \    {\rm for}\ \   g_\rho\sim 4.6\, .
\label{xiref}
\end{equation}
We could reduce the lower bound on $f$ to reach the Higgsless limit $\xi=1$, but
at the prize of having a very large $g_\rho$. In this case  the value of  $S$
can only be estimated, since it cannot be calculated by any perturbative method.  
In deriving Eq.~(\ref{xiref})  we have assumed that  $T$ receives a large and positive contribution, $\alpha \Delta T\sim 1-4\cdot 10^{-3}$, beyond that of the SM.
As we will see later, this can arise from one-loop  effects
that can be sizable if the top is  composite.

Finally,
in  the fermionic sector   we will take the following extra assumption.
The SM fermions   will be  assumed to  be linearly coupled to the BSM resonances.
This means that exists a basis  in which the SM fermions 
couple to the BSM sector only through mass mixing terms. 
In particular, for the top we have
\begin{equation}
\label{simplelagrangian}
\mathcal{L} = y_{L}f\, \bar{q}_{L}^{\rm el}\mathcal{P}_{q}[Q_{R}] + y_{R}f\,  \bar{t}_{R}^{\rm el}\mathcal{P}_{t}[T_{L}]+ M_{Q} \bar{Q}_{L}Q_{R}  + M_{T} \bar{T}_{R}T_{L} + g_{\rho} \bar{Q}_{L} \Sigma T_{R} + \cdots\, ,
\end{equation}
where ${q}_{L}^{\rm el}$ and ${t}_{R}^{\rm el}$ denote the elementary left-handed top-bottom doublet
and right-handed top respectively, and 
$Q_{L,R}$ and $T_{L,R}$ are  vector-like ``composite'' BSM resonances.
The operators   $\mathcal{P}_{q}$  and $\mathcal{P}_{t}$  project   the BSM resonances
into   components with the   SM quantum numbers of  $q_L^{\rm el}$  and $t_R^{\rm el}$ respectively.  We will   consider that there is only one $Q_{L,R}$ and $T_{L,R}$ resonance. 
In five-dimensional theories this corresponds to
keep only   the lightest   Kaluza-Klein (KK) state of each tower that it is usually
a good approximation \cite{Contino:2006nn}.
Apart from the mass terms,  we have included in Eq.~(\ref{simplelagrangian}) the  Yukawa term  $\bar{Q}_{L} \Sigma T_{R}$ responsible, as we will see, for the top mass.
The absence in Eq.~(\ref{simplelagrangian})
of  bilinear couplings of  elementary fields with  the BSM resonances, 
{\it e.g.} $\bar{q}^{\rm el}_{L} \Sigma t^{\rm el}_{R}$,  is a feature of holographic models
\cite{Agashe:2004rs}. It was also implemented in Technicolor models in 
Ref.~\cite{Kaplan:1991dc}.  
This implies that the top 
get a mass through mixing with  BSM states.
This way of generating fermion masses is phenomenologically favorable,
since it avoids dangerous flavor transitions  \cite{Agashe:2004rs} that were present in the original Technicolor models.
For our analysis here, however, the presence of terms like 
$\bar{q}^{\rm el}_{L} \Sigma t^{\rm el}_{R}$  would only introduce more parameters 
but would not qualitatively change our conclusions.

The SM top components, $q_L$  and $t_R$,  are identified with the massless
states (before EWSB). These are given by 
\begin{eqnarray}
\label{SMtop}
q_{L}&=&\cos{\theta_{L}}\,  q_L^{\rm el}+\sin{\theta_{L}}\, \mathcal{P}_{q}[Q_{L}]\ ,
\quad \tan{\theta_{L}} = \frac{y_{L}f}{M_{Q}}\, ,
\nonumber\\
t_{R}&=&\cos{\theta_{R}}\,  t_R^{\rm el}+\sin{\theta_{R}}\, \mathcal{P}_{t}[T_{R}]\ ,
\quad \tan{\theta_{R}} = \frac{y_{R}f}{M_{T}}\, .
\end{eqnarray}
The orthogonal states get a mass
squared $M^{2}_{Q}+y^{2}_{L}f^{2}$ and $M^{2}_{T}+y^{2}_{R}f^{2}$. 
The last term of   Eq.~(\ref{simplelagrangian})  gives, after the above rotation, the Yukawa coupling of the top:
\begin{equation}
\label{yukawatop}
y_{t} = g_{\rho} \sin{\theta_{L}}\sin{\theta_{R}}\, .
\end{equation}
By requiring  a top mass $m_t=y_t v\simeq 160$ GeV (at energies $M_\rho\sim 1$ TeV),
Eq.~(\ref{yukawatop})  
gives a lower bound   for the mixing angles, $\sin\theta_{L,R}  \gtrsim 0.6/g_\rho$.
The largeness of these mixing angles makes natural  the possibility  
that  one of the two chiralities of the top is fully  composite.  We will consider this possibility below. 

\subsection{The top composite limit}

We are  interested in exploring
the limit in which  either $q_L^{\rm el}$ or  $t_R^{\rm el}$ 
is maximally coupled to the BSM sector   
such that the SM $q_L$ or $t_R$  mostly 
corresponds to a composite BSM state. 
For the left-handed top,
this corresponds to the limit
\begin{equation}
\label{compolimitL}
\left\{  \begin{array}{c} \sin{\theta_{L}} \rightarrow 1 \\ y_{L} \rightarrow g_{\rho} \end{array}  \right\} \quad \textnormal{and} \quad \left\{  \begin{array}{c} \sin{\theta_{R}} \rightarrow y_t/g_\rho \\ y_{R} \simeq  y_t \end{array}  \right\}\, .
\end{equation}
For the right-handed top, the composite limit  is given by
\begin{equation}
\label{compolimitR}
\left\{  \begin{array}{c} \sin{\theta_{R}} \rightarrow 1 \\ y_{R} \rightarrow g_{\rho} \end{array}  \right\} \quad \textnormal{and} \quad \left\{  \begin{array}{c} \sin{\theta_{L}} \rightarrow y_t/g_\rho \\ y_{L} \simeq  y_t \end{array}  \right\}\, .
\end{equation}
In warped extra-dimensional   models these limits can be obtained by taking negative  values for the 5D  mass of the left-handed (or right-handed) top   that  localizes the 4D massless state towards the IR-boundary \cite{Agashe:2004rs}.
Although the composite limit can also be considered for  other SM fermions,
the fact that the top is the heaviest  of all of them  suggests that this is the most likely SM fermion    to have one of its chiralities being mostly composite.

Let us concentrate for the moment on the    $q_{L}$ composite  limit, Eq.~(\ref{compolimitL}).
In this limit  the SM left-handed top is part of the 
BSM  multiplet $Q_{L}$.   Since  $Q_{L}$ is in a  
SU(2)$_L\times$ SU(2)$_R$ representation,  
the top  will be accompanied  by custodial partners, the custodians, 
corresponding   to 
\begin{equation}
\label{project}
(1-\mathcal{P}_{q})[Q_{L}] \equiv \widetilde{\mathcal{P}}_{q}[Q_{L}] \, . 
\end{equation}
It is important to notice that the mass of the custodians is given by
$M_{Q}=y_Lf\cot\theta_L$ that in the composite limit  tends to zero.
Therefore in this  limit the custodian states   become lighter than the 
other resonances,    $M_Q\ll M_\rho$.
This effect has also been  observed in 5D  models 
in the limit in which the 5D masses take negative values and the massless states
become localized towards the IR-boundary \cite{Contino:2006qr}.
Nevertheless, 
it is hard to understand what could be the  origin of this new mass scale  $M_Q\ll M_\rho$
in a generic strongly-coupled theory.
The effect of having light custodians will have important phenomenological consequences
as we will see later.

Similarly,   in the  right-handed top  composite limit, Eq.~(\ref{compolimitR}),
one finds that the custodians, given by 
$(1-\mathcal{P}_{t})[T_{R}] \equiv \widetilde{\mathcal{P}}_{t}[T_{R}]$,
are also light $M_T\ll M_\rho$.

From now on we  will generically denote by $q^*$ the custodians and by $\mc$ their masses.

\section{Low-energy effective lagrangian for a composite top}
\label{effeclagra}

At energies below the resonance masses,
the effective theory corresponds to the SM plus higher-dimensional operators.
These operators are induced by integrating out the heavy resonances
at $M_\rho$ and the custodians at $\mc$.
In the first case, the higher-dimensional operators are suppressed  by $M_\rho$.
Among these operators, we will be interested in those carrying   extra
powers of  $g_\rho$ such that 
the effective scale that   
suppresses these operators is  in fact    $g_\rho/M_\rho=1/f$, 
that in the   limit considered here $g_\rho>1$, is larger than $1/M_\rho$.
These  are operators  with extra  composite tops 
or     Higgs fields  (or, in Higgsless theories,  the Goldstones)
which couple  to the BSM resonances with a coupling of order 
$g_\rho$.   
Let us present the list of these operators for the case of a  composite $q_L$,   Eq.~(\ref{compolimitL}). Up to order $p^2/f^2$,
we have three    dimension-6 operators  of this type \cite{Giudice:2007fh}
\begin{equation}
\label{effcompoL}
\frac{ic^{(1)}_{L}}{f^{2}}H^{\dagger}D_{\mu}H\bar{q}_{L}\gamma^{\mu}q_{L} 
+ \frac{ic^{(3)}_{L}}{2f^{2}}H^{\dagger}\sigma^{i}D_{\mu}H\bar{q}_{L}\gamma^{\mu}\sigma^{i}q_{L} + h.c. + \frac{c_{4q}}{f^{2}}(\bar{q}_{L}\gamma^{\mu}q_{L})(\bar{q}_{L}\gamma_{\mu}q_{L})\, .
\end{equation}
We are using the two-component notation $H$ for the Higgs  multiplet: 
\begin{equation}
\Sigma=(\widetilde H\, , H)\  \ \ \ \ {\rm where}\ \ \   H^\dagger H=v^2\, ,
\end{equation}
and $\widetilde H=i\sigma_2 H$.
Notice that we  are only including in $H$ the Goldstones  and not the  Higgs particle. 
The effects of  a composite Higgs  were already studied in Ref.~\cite{Giudice:2007fh}.
In the case where $v=f$,   we cannot 
expand in  $H/f$, and  we have, at the same leading order as the first two operators of 
Eq.~(\ref{effcompoL}), a  dimension-8  operator
\begin{equation}
\label{effcompoL2}
\frac{ic^{\prime}_{L}}{f^{4}}H^{\dagger}D_{\mu}H(\bar{q}_{L}H)\gamma^{\mu} (H^\dagger q_{L})\, .
\end{equation}
The second  class of operators that we  will be interested in  are those
induced by integrating out the custodians. These   operators are  suppressed by $\mc$. 
Since the $q_L$'s custodial partners  do not mix with  $q_L$
(they have different quantum numbers),   operators  induced at tree-level
cannot contain $q_L$.
The custodians of $q_L$, however,  can mix with $t_R$  through the Yukawa coupling
generating  higher-dimensional operators  involving $t_R$ and $H$ and 
carrying  powers of  $y_t^2/\mc^2$.
The leading operator of this kind is given by
\begin{equation}
\label{effeleR}
\frac{i\tilde c_{R}y^2_{t}}{\mc^{2}}H^{\dagger}D_{\mu}H\bar{t}_{R}\gamma^{\mu}t_{R}\, .
\end{equation}
At this point it is worth emphasizing   the crucial difference between the two classes of operators,  Eq.~(\ref{effcompoL}) and
Eq.~(\ref{effeleR}). 
The origin of the operator in  Eq.~(\ref{effeleR})  
is the mixing of $t_R$ with   the custodians. Therefore 
the strength of this   operator is related to the lightness of these extra states.
On the other hand, the strength of the operators 
in Eq.~(\ref{effcompoL}) measures the 
degree of compositeness of the top 
that do not have to be related to new light degrees of freedom.

We can repeat the same analysis for the case of  a composite $t_R$.
Up to order $ p^2/f^2$,  we have two operators \cite{Giudice:2007fh}
\begin{equation}
\label{effcompoR}
\frac{ic_{R}}{f^{2}}H^{\dagger}D_{\mu}H\bar{t}_{R}\gamma^{\mu}t_{R}+\frac{c_{4t}}{f^{2}}(\bar{t}_{R}\gamma^{\mu}t_{R})(\bar{t}_{R}\gamma_{\mu}t_{R})\, ,
\end{equation}
while  at order $p^2/\mc^2$   we have (from integrating out the  custodians of  $t_R$)
\begin{equation}
\label{effeleL}
\frac{i\tilde c^{(1)}_{L}y_t^2}{\mc^{2}}H^{\dagger}D_{\mu}H\bar{q}_{L}\gamma^{\mu}q_{L} + 
\frac{i\tilde c^{(3)}_{L}y^2_t}{2\mc^{2}}H^{\dagger}\sigma^{i}D_{\mu}H\bar{q}_{L}\gamma^{\mu}\sigma^{i}q_{L}+h.c.\, .
\end{equation}

The coefficients  $c_{i}$ are  $\mathcal{O}(1)$ constants whose values depend
on the details  of the BSM sector.
In certain cases, as we will see, these coefficients fulfill certain relations
due to the underlying symmetries of the BSM.
For a composite  Higgs model  the values of $c_{R,L}$ are given in Ref.~\cite{Agashe:2006at}.
In these models  the 
four-fermion interactions arise from integrating out  heavy vector resonances.
From a color resonance,  assuming a coupling $g_\rho$ to the top, one has
\begin{equation}
\label{c4tq}
c_{4t} =c_{4q} =-\frac{1}{6}\, ,
\end{equation} 
while for  a singlet resonance one gets $c_{4t} =c_{4q} =-1/2$.

\section{Present experimental constraints}
In this section we want to study  how much the present experimental data 
limits the compositeness of the top. Although
important effects  of the top compositeness   could  be revealed  in flavor physics,
we will not discuss   them here (see, however, Ref.~\cite{Giudice:2007fh}).
These effects 
strongly depend  on the underlying theory of flavor, and therefore  are very model dependent. Discarding flavor physics, the most stringent bound on the  composite $q_L$ case
comes from $Zb_L\bar b_L$ that has been measured at LEP at the per mille level.
This bound has strongly disfavored in the past  Technicolor models   and other variants \cite{Chivukula:1992ap}.
From the lagrangian of Eq.~(\ref{effcompoL}), we find
a deviation from the SM $Zb_L\bar b_L$  coupling given by 
\begin{equation}
\label{Zbbarb}
\frac{\delta g_{b_{L}}}{g_{b_{L}}} = \frac{(c_{L}^{(1)}+c_{L}^{(3)}) \xi}{1-\frac{2}{3}\sin^{2}{\theta_{W}}} \, .
\end{equation}
For $c^{(1),(3)}_L\sim 1$, as expected for a composite $q_L$,
Eq.~(\ref{Zbbarb})   gives  a  large deviation, excluded by the present LEP data.
This strong bound, however, can be evaded   in certain custodial BSM models. 
As pointed out in  Ref.~\cite{Agashe:2006at}, 
the custodial  symmetry implemented with  $P_{LR}$
(that interchanges $L\leftrightarrow R$)
can protect   $Zb\bar b$ from  large deviations from its SM value.
This occurs when the BSM field that couples  to   $b_L$ has
the following  isospin-left and isospin-right charge assignments \cite{Agashe:2006at}:
\begin{equation}
\label{ca}
T_L=T_R=1/2\ ,\ \  T^3_L=T^3_R=-1/2\, .
\end{equation}
In this case one finds, from integrating out the BSM sector,   $c_{L}^{(1)}= -c_{L}^{(3)}$, and 
therefore no contributions to Eq.~(\ref{Zbbarb}) are generated. 
The only effect on $Zb\bar b$ will arise from
loops involving SM particles (together with BSM states) that do not respect the custodial and $P_{LR}$ symmetry. We will comment on these effects later on.

Assuming  that Eq.~(\ref{ca}) is fulfilled, and that the operator $\bar Q_L\Sigma T_R$ 
must  be allowed to give masses to the SM fermions, we are left with 
only   two possible charge assignments for the states $Q$ and $T$ under SU(2)$_L\times$SU(2)$_R\times$U(1)$_X$ \footnote{The extra global U(1)$_X$ symmetry of the BSM sector is needed to properly  embed   the hypercharge of the SM, $Y=T^3_R+X$.}:
\begin{equation}
\begin{array}{|c|c|c|}\hline & Q & T \\\hline { \text{Case\ (a)}} & {\bf (2,2)_{2/3}} & {\bf (1,1)_{2/3}} \\\hline { \text{Case\ (b)}} & {\bf (2,2)_{2/3}} & {\bf (1,3)_{2/3}+(3,1)_{2/3}}\\\hline \end{array}
\label{cases}
\end{equation}
In this article we will concentrate only on these two possibilities.

\subsection{The $\widehat T$ parameter}
With $Zb\bar b$ under control at tree-level,  the next important observable is the 
$T$-parameter.
The contribution to $T$  arises from the higher-dimensional operator
\begin{equation}
\frac{c_T}{2f^2}|H^\dagger D_\mu H|^2\ ,\quad \widehat T=c_T\xi\, ,
\end{equation}
where  we follow the notation  of Ref.~\cite{Barbieri:2004qk} in which  the $T$-parameter is rescaled: 
$\widehat T=\alpha T\simeq T/129$. 
As we previously said, 
$\widehat T$  is zero at the tree-level by the custodial symmetry. Nevertheless,  it can be generated 
at the one-loop level due to the $y_{L,R}$ couplings in Eq.~(\ref{simplelagrangian}) which break the custodial symmetry.
A dimensional  estimate shows that \cite{Giudice:2007fh} 
\begin{equation}
\widehat T\sim  \frac{N_c}{16\pi^2}\left(\frac{ y_{L,R}}{g_\rho}\right)^4\frac{\xi\Lambda^2}{f^2}\, ,
\label{Testimate}
\end{equation}
where $N_c=3$ is the QCD number of colors 
and $\Lambda$ is the cutoff scale. If  $\Lambda\sim M_\rho$
we get a very large   contribution, forbidding 
the composite region $y_{L,R}\sim g_\rho$.
Nevertheless, we must recall that in the top composite limit, the custodians are light
$\mc< M_\rho$, and, as we will see,  are their masses  what really cut off the loop momentum.
Therefore we cannot neglect the effects of the custodians $Q$ and $T$
that can  diminish the bound on $y_{L,R}$ and allow
a higher   degree of compositeness for the top. 

\begin{figure}[t!]
	\begin{center}
	\includegraphics[width=4.3in]{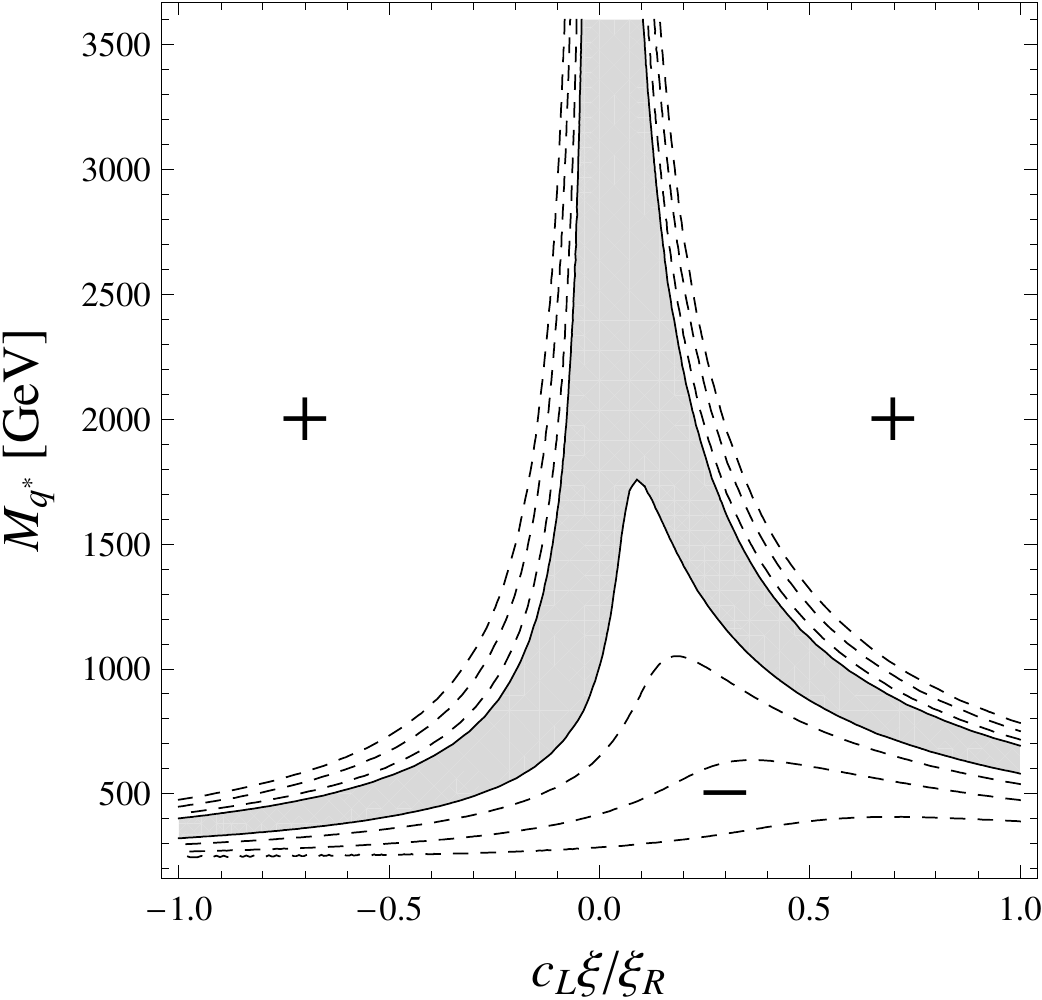}
	\caption{
Contribution to 	$\widehat{T}$  in the  $q_L$ composite  limit (case (a))
in the $\mc-c_{L}\xi/\xi_{R}$ plane,  where $\xi_{R}=1/4$.
The grey area shows the  region $-1.7\cdot 10^{-3}<\widehat T<+1.9\cdot 10^{-3}$ and  the dashed lines show the 
contribution to $|\widehat T|$ equal to   2.8, 4.2 and 5.6 as they respectively move away
from the grey area.  We have 
marked  with a  ``$+$" (``$-$") the areas in which the  contribution to $\widehat T$ is positive (negative). The dotted line corresponds to the holographic composite Higgs model.
	}
	\label{T11}
	\end{center}
\end{figure}

\begin{figure}[t!]
	\begin{center}
	\includegraphics[width=3.3in]{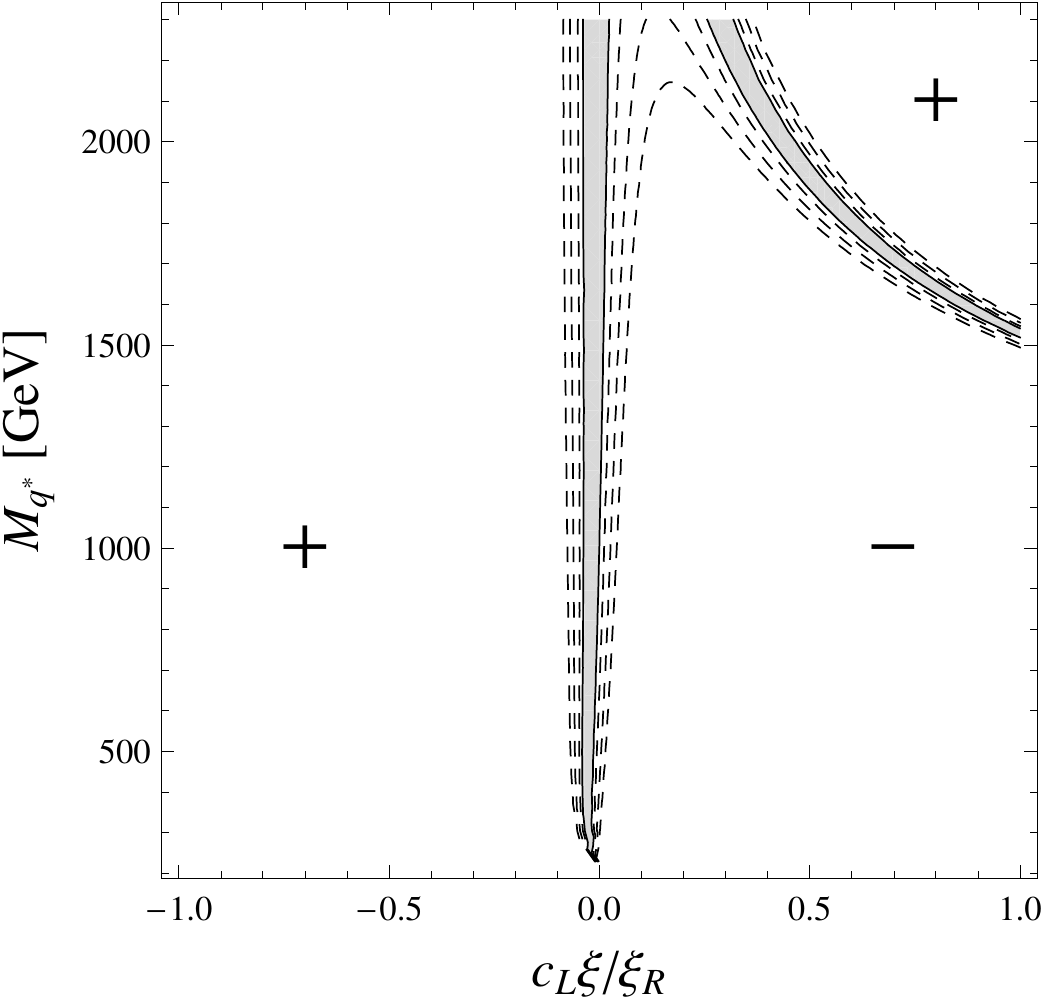}
	\includegraphics[width=3.3in]{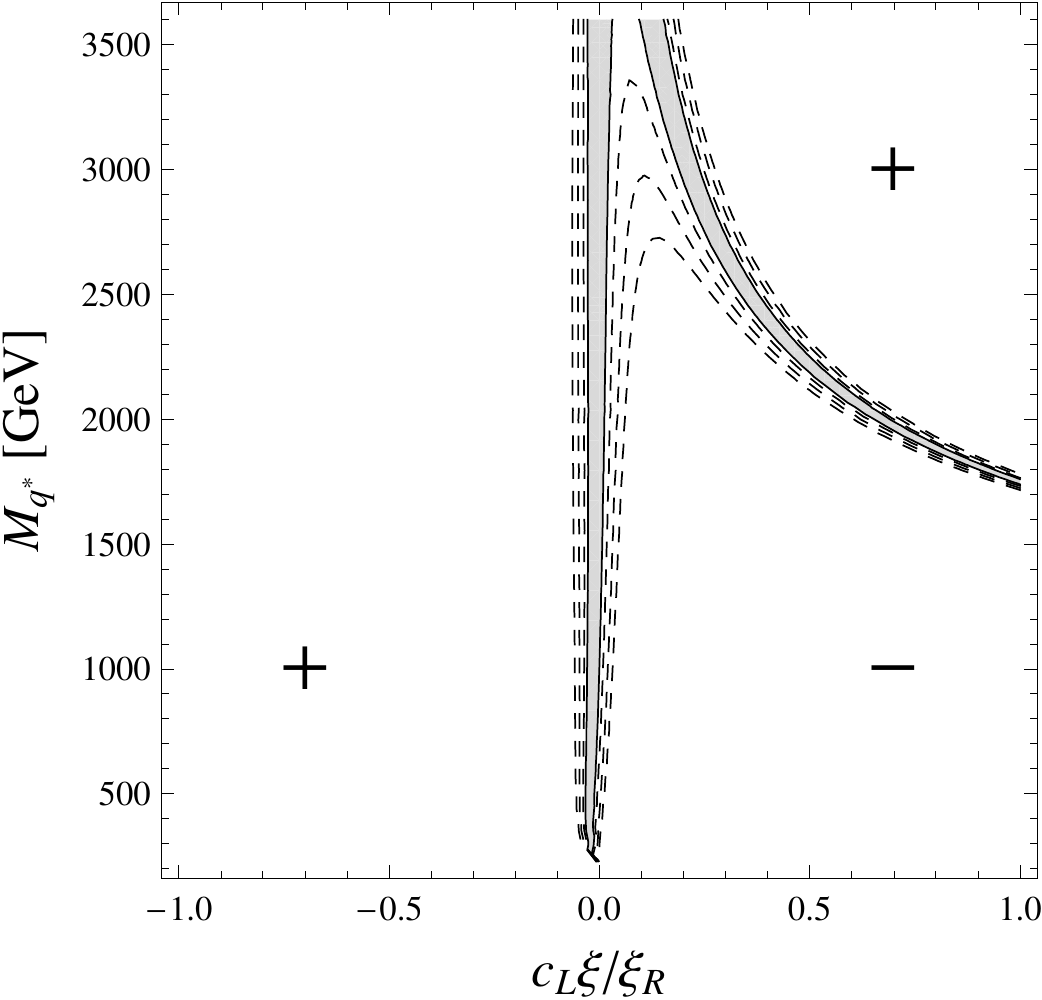}
	\caption{
	Contribution to 	$\widehat{T}$ in the  $q_L$ composite  limit (case (b))
in the $\mc-c_{L}\xi/\xi_{R}$ plane,  where $\xi_{R}=1/4$.
The grey area shows the  region $-1.7\cdot 10^{-3}<\widehat T<+1.9\cdot 10^{-3}$ and  the dashed lines show the 
contribution to $|\widehat T|$ equal to   2.8, 4.2 and 5.6 as they respectively move away
from the grey area.  We have 
marked  with a  ``$+$" (``$-$") the areas in which the  contribution to $\widehat T$ is positive (negative). 
We have taken $M_{\rho}=2.3$ TeV (left) and  $M_{\rho}=3.6$ TeV (right). 	}
	\label{T13}
	\end{center}
\end{figure}

\begin{figure}[t!]
	\begin{center}
	\includegraphics[width=3.3in]{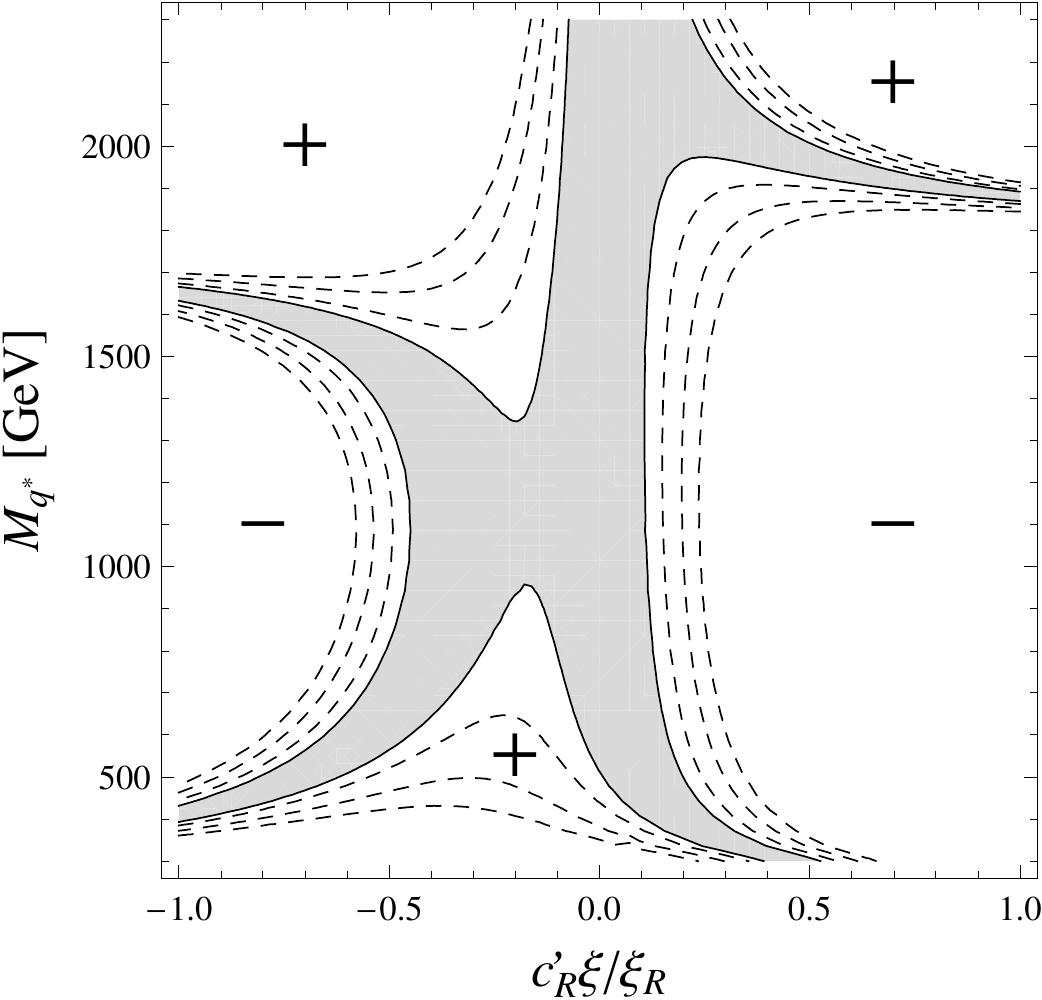}
	\includegraphics[width=3.3in]{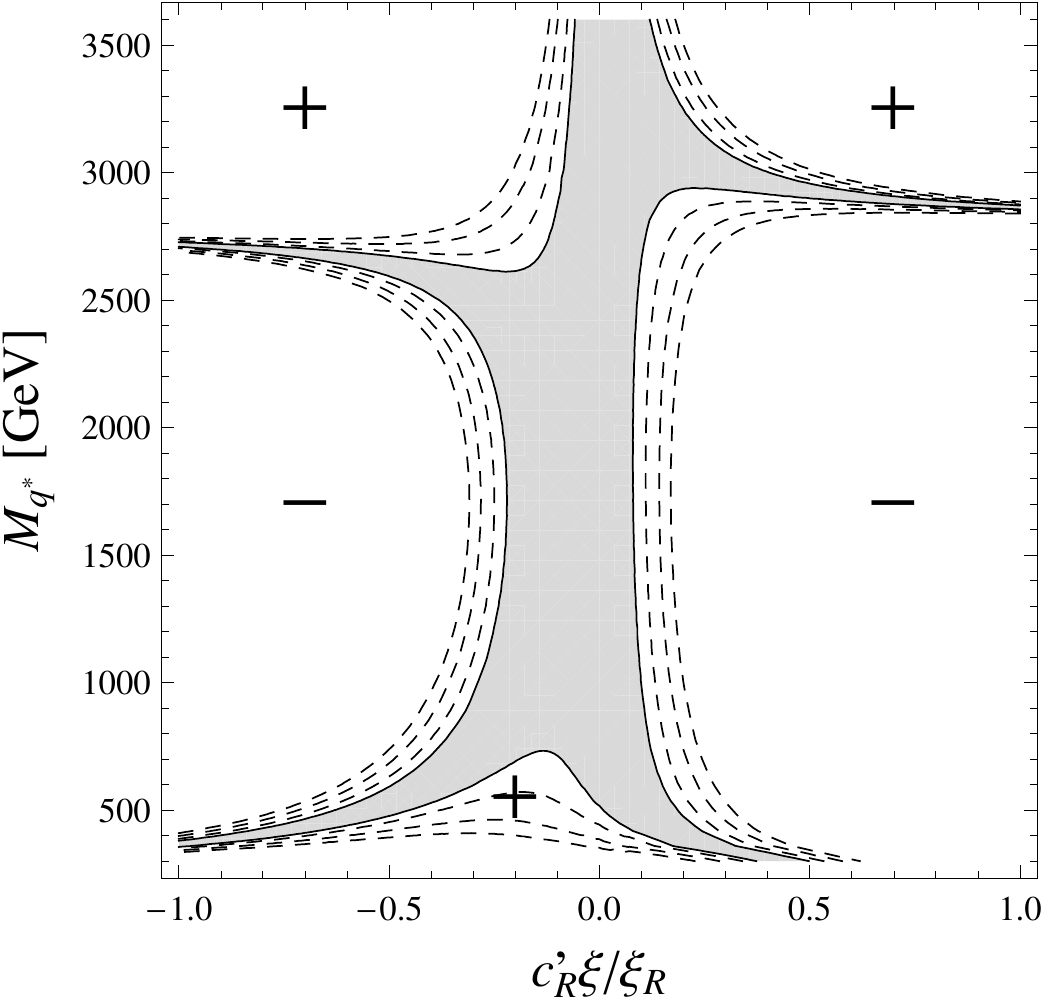}
	\caption{
Contribution to 	$\widehat{T}$  in the  $t_R$ composite  limit (case (b))
in the $\mc-c_{R}^\prime\xi/\xi_{R}$ plane,  where $\xi_{R}=1/4$.
The grey area shows the  region $-1.7\cdot 10^{-3}<\widehat T<+1.9\cdot 10^{-3}$ and  the dashed lines show the 
contribution to $|\widehat T|$ equal to   2.8, 4.2 and 5.6 as they respectively move away
from the grey area.  We have 
marked  with a  ``$+$" (``$-$") the areas in which the  contribution to $\widehat T$ is positive (negative).
We have taken $M_{\rho}=2.3$ TeV (left) and  $M_{\rho}=3.6$ TeV (right). }
	\label{T1331}
	\end{center}
\end{figure}

We have performed the calculation of $\widehat T$  in the  $q_L$ and $t_R$ composite
limits  taking into account the custodians.
We have considered  the two charge assignments (a) and (b) of Eq.~(\ref{cases}).
For a composite $q_L$ 
the  results of $\widehat T$ are plotted in Figs.~\ref{T11} and  \ref{T13}
for the charge assignment (a) and (b) respectively.
They depend on the mass of the custodians, $\mc$, and 
the  coefficient of the higher-dimensional operator
$c_L\equiv c^{(3)}_L=- c^{(1)}_L$.
For a  composite $t_R$, only the charge assignment (b) 
gives a nonzero contribution to $\widehat T$.
This is plotted in  Fig.~\ref{T1331}.
In this case the constraints on $\widehat T$  do not give any direct bound
on the coefficients $c_i$ of Eq.~(\ref{effcompoR}),
but only on the coefficient of the higher-dimensional operators of the custodians $c^\prime_{R}$. 

To  understand these results we will present  the
calculation of $\widehat T$  in the limit $M_\rho\gg \mc\gg m_t$
following the effective theory approach of Ref.~\cite{Cohen:1983fj}. 
This consists in calculating the leading effects to $c_T(\mu)$ 
at the three different values of the renormalization scale $\mu$:
At $\mc<\mu<M_\rho$ in the effective theory after integrating out the heavy resonances, at $m_t<\mu<\mc$  after integrating out the custodians,
and finally at $\mu< m_t$ after integrating out the top.

Let us start with the  $q_L$ composite  limit:

\noindent{\bf Case (a):}  
The theory below  $M_\rho$ but above $\mc$ 
consists of  the SM plus the custodians. 
The $q_L$ and its custodians $q^*_L$ are embedded 
in the $(\textbf{2},\textbf{2})_{\textbf{2/3}}$ representation denoted by $Q_L$.
Under the SM SU(2)$_L\times$U(1)$_Y$ group, $q^*_L$  transforms as  a $\bf 2_{7/6}$. 
We choose to represent $Q_{L}$ by a $2 \times 2$ matrix given by $Q_{L}=(q_{L}, q^{*}_{L})$. 
The dimension-4 operators involving the top  and the custodians are given by
\begin{equation}
{\cal L}_4=\Tr[\bar{Q}_{L} i \gslash{D} Q_{L}] +
\Tr[\bar{Q}_{R} 
i  \gslash{D} 
Q_{R}\widetilde{P}_{q}]
+ \bar{t}_{R} i \gslash{D} t_{R}+\Big\{ y_{t} \Tr[\bar{Q}_{L}\Sigma P^{-1}_{t}] t_{R} +
\mc \Tr[ \bar{Q}_{L} Q_{R} \widetilde{P}_{q}] + h.c.\Big\}\, ,
\label{ope4}
\end{equation}
where $P_{t}^{-1} = \mathbb{I}$ follows from the embedding $T_{R} \equiv (\mathbf{1},\mathbf{1})$  and $\widetilde{P}_{q} = (1-\sigma_{3})/2$.
Notice that the  only breaking of the custodial symmetry arises from the custodian mass
term due to the presence of $\widetilde{P}_{q}$.
There are also  dimension-6 operators  that can contribute to $\widehat T$.
Up to  order  $p^2/f^2$, they are given by
\begin{equation}
{\cal L}_6=\frac{c_{L}}{f^{2}}\Big\{\textnormal{Tr}[\bar{Q}_{L}\gamma^{\mu}Q_{L}\hat{V}_{\mu}] +\textnormal{Tr}[\bar{Q}_{L}\gamma^{\mu}V_{\mu}Q_{L}]\Big\},
\label{ope6}
\end{equation}
where $c_L$ is a coefficient of order one and we have defined $V_{\mu}=(iD_{\mu}\Sigma)\Sigma^{\dagger}$, $\hat{V}_{\mu} = (iD_{\mu}\Sigma)^{\dagger}\Sigma$, and the covariant derivative is given by $D_{\mu}\Sigma = \partial_{\mu}\Sigma - ig\sigma_{a}W^{a}_{\mu}\Sigma/2 + ig'B_{\mu}\Sigma\sigma_{3}/2$. We are omitting 
the  double-trace operator $\textnormal{Tr}[\bar{Q}_{L} i \gslash{D} \Sigma]\textnormal{Tr}[\Sigma^{\dagger}Q_{L}]$ since this is 
suppressed in 5D theories \cite{Agashe:2006at} or
strongly-coupled theories in the large-$N$ limit.
The fact that the   two operators in Eq.~(\ref{ope6}) have equal coefficients is a consequence
of the   $P_{LR}$ symmetry.
We are neglecting operators suppressed by $\mc^2/M_\rho^2$
that we consider small in the top composite   limit.

At  the order that we are working,
the coefficient $c_T$ does not receive any contribution 
from integrating out the resonances at $M_\rho$ 
\footnote{We are not considering the contribution coming from a loop 
of gauge bosons.}.
To see this,   notice that the one-loop contribution  to $\widehat T$
arising from the effective  lagrangian 
Eqs.~(\ref{ope4}) and (\ref{ope6}) is  finite, {\it i.e.},  insensitive to the cutoff $M_\rho$.
This is a consequence of the custodial symmetry.
Indeed, the parameter   $\widehat T$, that  transforms as a $\bf 5$ under 
the custodial SU(2)$_V$ \cite{Kennedy:1991wa}, can only be 
generated  
from   diagrams with  at least four $\mc$ insertions, since
$\mc$  transforms as a   $\bf 2$ under SU(2)$_V$ (as a   ${\bf (1,2)}$  under SU(2)$_L\times$SU(2)$_R$).
This renders the custodian loop diagrams to $\widehat T$  finite \footnote{This does
not mean that the custodian contribution to $\widehat T$  must be proportional
to $\mc^4$. 
Diagrams with  four $\mc$ insertions contributing to  
$\widehat T$ are
UV-finite but  infrared divergent $\widehat T\propto \mc^4/\Lambda_{IR}^2$.
The infrared divergence is cure by the same $\mc$  when resumming 
over all possible  $\mc$ insertions, 
giving a final contribution  $\widehat T \propto \mc^2$.
Similar argument explains the finiteness  of the SM top  contribution to $\widehat T$
and its proportionality  to  $m_t^2$.}.
Our explicit calculation below will confirm this expectation.

Let us  now integrate out the custodians. Apart from  SM terms,
this  generates  the effective lagrangian terms
of Eqs.~(\ref{effcompoL}) and (\ref{effeleR})  with the coefficients 
\begin{equation}
\label{matchL}
c_{L}^{(3)} =-c_{L}^{(1)} = c_{L}\ ,\quad c_L^\prime=0\ ,\quad\tilde c_{R} = 1.
\end{equation}
To obtain  the coefficient $c_T$ at the custodian mass scale we must 
use the  matching condition at this boundary $\mu=\mc$ which is given by
\begin{equation}
\label{match1}
\widehat{T}_{total} = \widehat{T}_{custodians}+\widehat{T}_{top}+\widehat{T}_{mix} = c_{T}(\mc)\xi + \widehat{T}_{top},
\end{equation}
where $\widehat{T}_{total}$ includes the contributions from all the scales to the $\widehat{T}$ parameter, and   $\widehat{T}_{custodians}$, $\widehat{T}_{top}$ and $\widehat{T}_{mix}$ includes respectively   those arising from loops of   custodians,   tops and both. 
$\widehat{T}_{top}$  drops   in Eq.~(\ref{match1}) since we are not yet integrating out 
the top. The three contributions, $\widehat{T}_{custodians}$, $\widehat{T}_{top}$ and $\widehat{T}_{mix}$ separately, are understood as being renormalized in the $\overline{MS}$ scheme. Therefore, our matching condition for $c_{T}$ becomes 
\begin{equation}
\label{cTmS}
\xi c_{T}(\mc) = \widehat{T}_{top}^{SM} \left( 2 c_{L}^{2} \frac{\xi^{2}}{\epsilon_{t}} + 6 c_{L}^{2}\xi^{2} + 8 c_{L}\xi + \frac{22}{3}\epsilon_{t} \right)\, ,
\end{equation}
where we have kept the leading and subleading terms   in the expansion  parameter
\begin{equation}
\label{epst}
\epsilon_{t} = \frac{m_{t}^{2}}{\mc^{2}}\ll 1,
\end{equation}
and we have defined $\widehat{T}_{top}^{SM}$ as
\begin{equation}
\label{tsm}
\widehat{T}_{top}^{SM} =\frac{3m^2_t}{16\pi^2v^2}\simeq 0.008\, ,
\end{equation}
that  is equal to the  SM-top leading-contribution    to $\widehat T$.
It is important to note that all except the first term in the $\textnormal{r.h.s.}$ of Eq.~(\ref{cTmS}) are scheme-dependent. 
This first term shows that, as expected,   the quadratic divergence scale of 
Eq.~(\ref{Testimate}) is replaced  by $\mc^2$.

\begin{figure}[t!]
\begin{center}
		\includegraphics[width=6.5in]{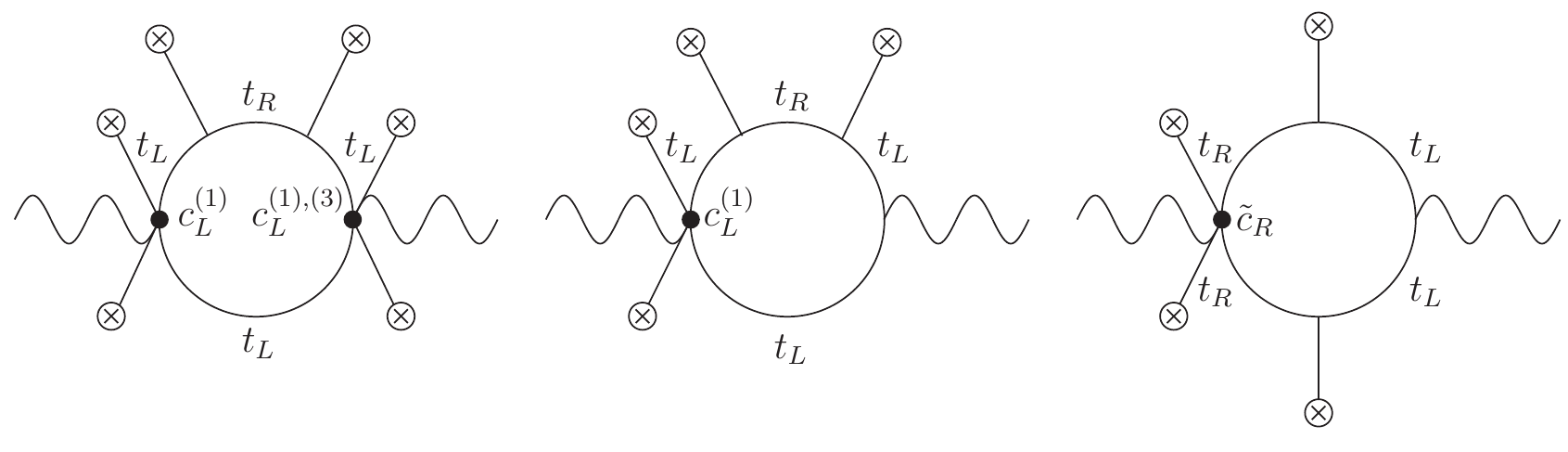}
\caption{Logarithmic divergent loop diagrams contributing to the SM gauge boson masses,
and therefore to $c_{T}$, for the low-energy effective theory  of a composite $q_L$,
Eqs.~(\ref{effcompoL}) and (\ref{effeleR}). The external  lines  with a cross correspond to  insertions of the Higgs  VEV.}
\label{log}
\end{center}
\end{figure}
Now,  we must use the renormalization group to scale $c_{T}(\mc)$ down to the lower scale $m_{t}$, where we can integrate out the top quark. 
The leading  logarithmic terms arise from the diagrams of Fig.~\ref{log}.
We obtain the equation
\begin{equation}
\label{rellog}
\xi c_{T}(m_{t}) = \xi c_{T}(\mc) + \widehat{T}_{top}^{SM} \left( 6 c_{L}^{2} \xi^{2} + 4 c_{L}\xi + 4\tilde  c_R\epsilon_{t} \right) \log \epsilon_{t}\, .
\end{equation}
Finally,  we must integrate out the top. The
matching condition at the boundary $\mu = m_{t}$ is given by
\begin{equation}
\label{match2}
\left[ \xi c_{T}(\mu) + \widehat{T}_{top} \right]_{\mu \rightarrow m_{t}^{+}} = \left[ \xi c_{T}(\mu) \right]_{\mu \rightarrow m_{t}^{-}}\, ,
\end{equation}
where  in the $\overline{MS}$ scheme 
\begin{equation}
\label{final}
\left[ \widehat{T}_{top} \right]_{\mu \rightarrow m_{t}^{+}} = \widehat{T}_{top}^{SM} (c_{L}^{2}\xi^{2} + 2 c_{L}\xi)\, .
\end{equation}
Here we are not including the SM top contribution to $c_T$ since we want
only the  contribution  to $\widehat T$ beyond the one of the SM.
Adding up    Eqs.~(\ref{cTmS}), (\ref{rellog}), and (\ref{final}) we obtain
\begin{equation}
\label{Ttotal}
\widehat{T}=\xi c_T(0)=
\widehat{T}_{top}^{SM} \left[ c_{L}^{2}\xi^{2} \left( \frac{2}{\epsilon_{t}} + 7 + 6 \log \epsilon_{t} \right) + c_{L}\xi \left( 10 + 4 \log \epsilon_{t} \right) + \epsilon_{t} \left( \frac{22}{3} + 4 \log \epsilon_{t} \right) \right]\, .
\end{equation}
As explained before, this result is valid in the limit $M_{\rho} \gg \mc \gg m_{t}$. We have checked that in this limit it agrees  with the  exact calculation.

In Fig.~\ref{T11} we present a plot of  $\widehat{T}$ (the exact result)
in the $\mc-c_{L}\xi/\xi_{R}$ plane, where $\xi_{R}=1/4$ is the reference value 
of $\xi$ in composite Higgs models --see Eq.~(\ref{xiref}).
The grey area shows the  region $-1.7\cdot 10^{-3}<\widehat T<+1.9\cdot 10^{-3}$ and  the dashed lines show the 
contribution to $|\widehat T|$ equal to   2.8, 4.2 and 5.6 as they respectively move away
from the grey area; we have 
marked  with a  ``$+$" (``$-$") the areas in which the  contribution to $\widehat T$ is positive (negative).
We see that the region of a  composite top,   $c_L\xi/\xi_R\sim 1$, is allowed
although,   as we expected,   requires  light custodians $\mc \lesssim 1$ TeV.
This correlation between $\mc$ and $c_L$  
tells us that the custodians must be seen at the LHC if $q_L$ is a fully composite state. 
Fig.~\ref{T11} also shows the region in which 
$\widehat T$ gets a positive contribution,
as needed in composite Higgs or Higgsless models in order to satisfy  EWPT.
We see that a positive contribution  $\widehat T\sim 1-4\cdot 10^{-3}$
is easily achieved for a composite top,  especially for  negative  values of $c_L$ and large values of the custodian
mass.
For small values of $\mc$, we obtain however a   negative value for $\widehat T$ 
that can be easily understood as follows.
In the  lagrangian Eqs.~(\ref{ope4}) and (\ref{ope6})  
the scale $\mc$ is the only  breaking parameter of the custodial symmetry.
Therefore in the limit $ \mc \rightarrow 0$ 
we must get  that the total contribution of the top and custodian sector must be zero,
implying  that the custodian contribution is given by
$\widehat{T} = -\widehat{T}^{SM}_{top}<0$.
In Fig.~\ref{T11}
we also show, with  a dotted line, the prediction for the   holographic Higgs model
\cite {Contino:2006qr}
in which  $\xi\sim 1/4$ and $\mc\sim 2.3\sqrt{1-2 c_L}$ TeV.
Notice that in this model  the contribution to $\widehat T$ is   negative, as it is also 
shown in Ref.~\cite{Carena:2007ua}.

\begin{figure}[t!]
\begin{center}
	\includegraphics[width=4.7in]{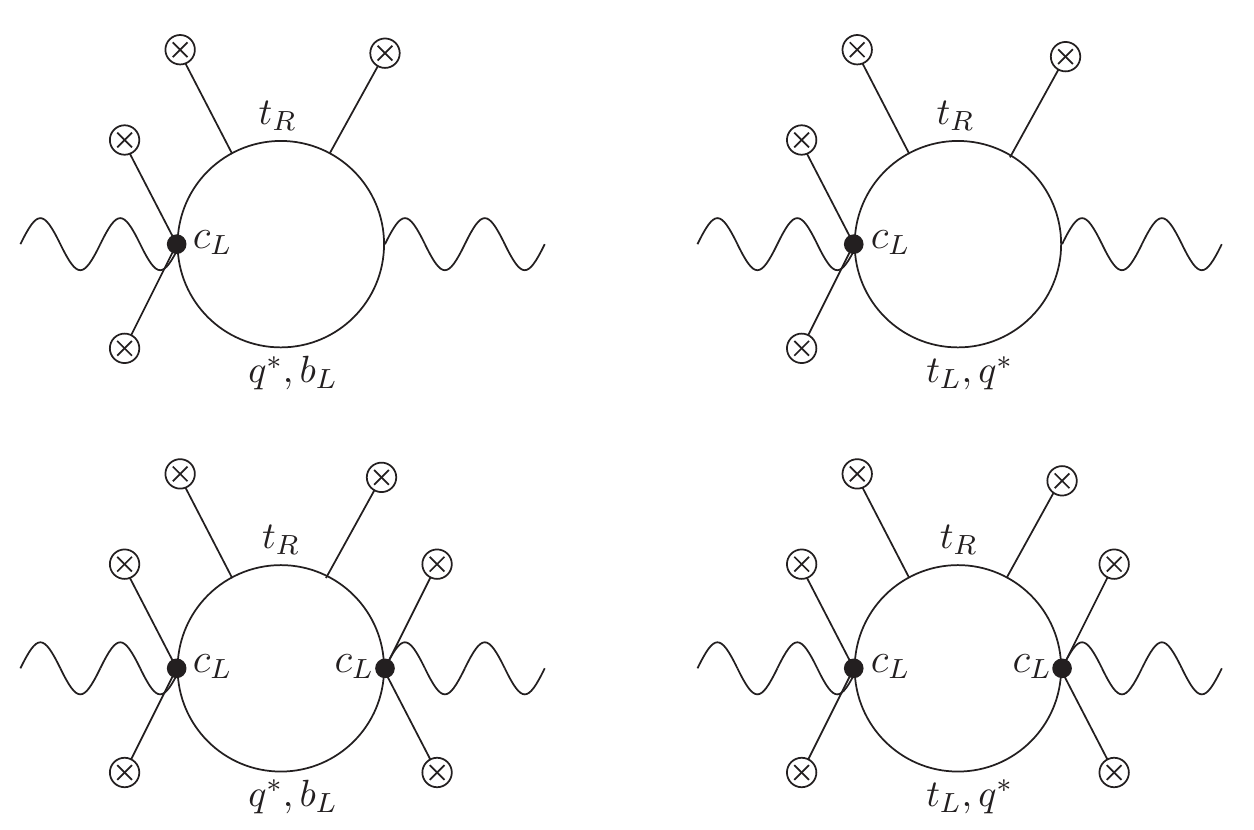}
\caption{Logarithmic divergent loop diagrams contributing to  
the SM gauge boson masses,
and therefore to $c_{T}$, for the low-energy effective theory of a composite $q_L$ and 
its custodians (case (b)).
The external  lines  with a cross correspond to  insertions of the Higgs  VEV.}
\label{logmrho}
\end{center}
\end{figure}

\noindent{\bf Case (b):}
In this case the representation of $T_R$  is
${\bf (1,3)_{2/3}+(3,1)_{2/3}}$
that implies
that the low-energy effective lagrangian for the top and the custodians
below $M_\rho$
is the same as that of 
Eqs.~(\ref{ope4})
and (\ref{ope6}) but with    $P_{t}^{-1} = \sigma_{3}$.
Now the  breaking of the  custodial symmetry not only  comes from the custodian mass term but also from the Yukawa coupling \footnote{This latter breaking arises from the fact that $y_R\simeq y_t$ in Eq.~(\ref{simplelagrangian}) breaks  the custodial symmetry.}.
This implies  that, contrary to the case (a),
the one-loop contribution to $c_T$ is not finite.
Indeed, the Yukawa coupling transforms as a $\bf 3$ under  the custodial symmetry 
SU(2)$_V$,
and therefore contributions to $\widehat T$ (a $\bf 5$ of SU(2)$_V$) only need    two powers of $y_t$.
In this case, as shown in Fig.~\ref{logmrho},    there are  custodian diagrams  
contributing to $\widehat T$ that  are   logarithmically UV-divergent. 

We have  now  
$c_T(M_\rho)\propto y_t^2$  that, being   sensitive to the physics at $M_\rho$, cannot
be predicted within our effective lagrangian approach.  What it is calculable, however, is the evolution of the coefficient $c_T(\mu)$
from $\mu=M_\rho$  to $\mu=\mc$ that comes from  the diagrams of  Fig.~\ref{logmrho}.
We obtain
\begin{equation}
\label{runnnn}
\xi c_{T}(\mc) = \xi c_{T}(M_{\rho}) - \widehat{T}_{top}^{SM} 16 \left( c_{L}^{2}\xi^{2} + c_{L}\xi \right) \log \left( M_{\rho}^{2}/\mc^{2} \right).
\end{equation}
From now on  we will define   $M_\rho$    by the scale at which $c_{T}(M_{\rho})=0$.
Let us now integrate out the custodians. The
coefficients of the  effective lagrangian of the top are the same as those in Eq.~(\ref{matchL}).
For $c_T$, the matching condition at $\mu = \mc$ reads
\begin{equation}
\label{match3}
\left[ \xi c_{T}(\mu) \right]_{\mu \rightarrow \mc^{-}} = \left[ \xi c_{T}(\mu) + \widehat{T}_{custodians} + \widehat{T}_{mix} \right]_{\mu \rightarrow \mc^{+}}\, ,
\end{equation}
where
\begin{equation}
\label{resss}
\left[ \widehat{T}_{custodians} + \widehat{T}_{mix} \right]_{\mu = \mc^{+}} = \widehat{T}_{top}^{SM} \left( 2 c_{L}^{2} \frac{\xi^{2}}{\epsilon_{t}} + 6c_{L}^{2}\xi^{2} - 8 c_{L}\xi + \frac{22}{3}\epsilon_{t} \right)\, .
\end{equation}
Including the evolution of $c_T$ from $\mc$ to $m_t$ and 
integrating out the top, that proceeds exactly as in the previous case, we end up with
\begin{eqnarray}
\label{Ttotal3}
\widehat{T}&=& \widehat{T}_{top}^{SM}\left[ c_{L}^{2}\xi^{2} \left( \frac{2}{\epsilon_{t}} + 7 + 6 \log \epsilon_{t} - 16 \log \frac{M_{\rho}^{2}}{\mc^{2}} \right)\right.\nonumber\\
&+& \left.c_{L}\xi \left( -6 + 4 \log \epsilon_{t} - 16 \log \frac{M_{\rho}^{2}}{\mc^{2}} \right)
+ \epsilon_{t} \left( \frac{22}{3} + 4 \log \epsilon_{t} \right)\right]\, . 
\end{eqnarray}
The exact value of $\widehat{T}$ in the $\mc-c_{L}\xi/\xi_{R}$ plane is presented 
in Fig.~\ref{T13} for  $M_{\rho}\simeq 2.3$ TeV (left)
and $M_\rho\simeq 3.6$ TeV (right). The region of sizable  values of 
$c_{L}\xi/\xi_{R}$ is extremely reduced due to the logarithms of Eq.~(\ref{runnnn}), disfavoring the possibility of a composite $q_L$  in this case.
This analysis, however, is useful to show that  regions 
with  positive contributions to $\widehat{T}$ are quite generic;
they  correspond to  $c_L<0$.
Since previous studies of the effects of
$\widehat{T}$ \cite{Carena:2007ua}
centered  in minimal holographic models in which  $c_L>0$, 
these regions with positive $\widehat T$ were overlooked.

Let us now consider the  $t_R$ composite  limit:

\noindent{\bf Case (a):} 
In this case $T_R$ is a singlet that corresponds, in the limit Eq.~(\ref{compolimitR}),  to $t_R$. There are no custodians
and the effective theory below $M_\rho$ corresponds to the SM plus the 
operators of  Eqs.~(\ref{effcompoR}).
We find
\begin{equation}
\label{coefftrcompo}
c_R=0\,   ,
\end{equation}
that is a consequence  of the custodial symmetry \cite{Agashe:2006at}. 
Eq.~(\ref{coefftrcompo}) together with  the absence of custodians imply that 
$\widehat T$ is not generated at the order considered here.
Hence, no serious bounds  on  a composite $t_R$ are obtained  in this case.

\begin{figure}[t!]
\begin{center}
	\includegraphics[width=4.7in]{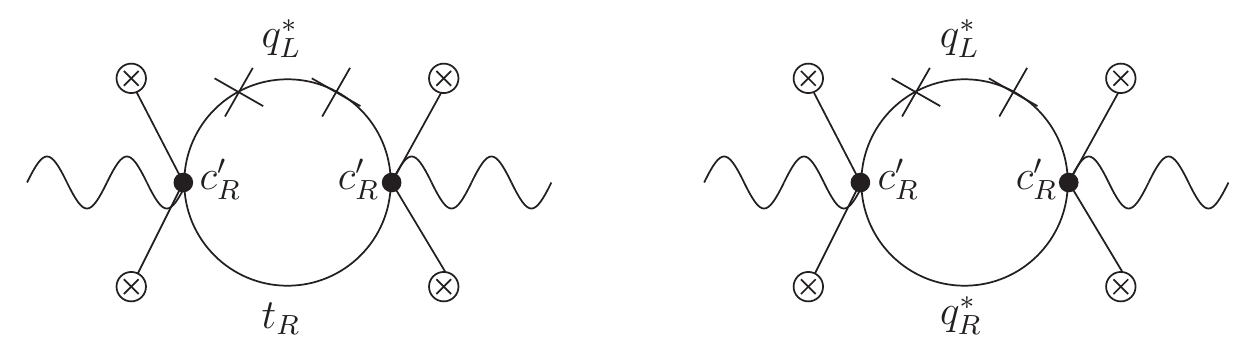}
\caption{Logarithmic divergent loop diagrams contributing to the SM gauge boson masses,
and therefore to $c_{T}$, for the low-energy effective theory  of a composite $t_R$
and its custodial partners (case (b)).  The external  lines  with a cross correspond to  insertions of the Higgs  VEV and the crosses denotes $\mc$ insertions.
}
\label{logmrhotR}
\end{center}
\end{figure}

\noindent{\bf Case (b):}
In this case $t_R\in T_{R}^{(1)}+T_{R}^{(2)}$ transforming  as a $(\mathbf{1},\mathbf{3})_{\mathbf{2/3}} + (\mathbf{3},\mathbf{1})_{\mathbf{2/3}}$. 
There are then  five custodians that  transform as $\bf 1_{5/3}$, $\bf 1_{-1/3}$ and $\bf 3_{2/3}$ under the electroweak symmetry.  
Using a $2 \times 2$ matrix representation for $T_{R}^{(1),(2)}$, we have the following
dimension-4 operators for the  top and custodians: 
\begin{eqnarray}
\mathcal{L}_{4} & =& \Tr[\bar{T}_{R}^{(1)} i \gslash{D} T_{R}^{(1)}] + \Tr[\bar{T}_{R}^{(2)} i \gslash{D} T_{R}^{(2)}] + \Tr[\bar{T}_{L}^{(1)} i \gslash{D} T_{L}^{(1)}] + \Tr[\bar{T}_{L}^{(2)} i \gslash{D} T_{L}^{(2)}] + \bar{q}_{L} i \gslash{D} q_{L}\nonumber\\
&+ & y_{t}\sqrt{2} \Tr[(\bar{T}_{R}^{(1)}
\Sigma^{\dagger}+
\Sigma^{\dagger}\bar{T}_{R}^{(2)}) \mathcal{P}_{q}^{-1}(q_{L})] 
+ \mc\left\{ \Tr[\bar{T}_{R}^{(1)} \widetilde{P}_{t} T_{L}^{(1)}] +\Tr[ \bar{T}_{R}^{(2)}T_{L}^{(2)}]\right\} + h.c.\, ,
\label{LMtc}
\end{eqnarray}
where $\mathcal{P}_{q}^{-1}(q_{L}) =(q_L,0)$  and $\widetilde{P}_{t}=\sigma_{3}$.
These two projectors, appearing in the Yukawa and custodian masses,
parametrize the breaking of the custodial symmetry. 
Contributing to $\widehat{T}$, there can also be 
dimension-6 operators that, 
up to order $p^2/f^2$, are given by     
\begin{equation}
\label{rightcusto}
\frac{c^\prime_{R}}{f^{2}}\left\{\Tr\left[\bar{T}^{(1)}_{R}\gamma^{\mu}[\hat{V}_{\mu},T^{(1)}_{R}]\right]-\Tr\left[\bar{T}^{(2)}_{R}\gamma^{\mu}[V_{\mu},T^{(2)}_{R}]\right]\right\}\, .
\end{equation}
The contribution of the above lagrangian to $c_T$ is  logarithmically divergent
\footnote{We can see this by assigning to $y_t$ and $\mc$ the  representation $\bf (1,2)$  and $\bf(1,3)$ respectively   to make the lagrangian SU(2)$_L\times$SU(2)$_R$ invariant.
Therefore $\widehat T$ must arise from diagrams with four powers of $y_t$  and two of $\mc$. The diagrams with two $\mc$ insertions (Fig.~\ref{logmrhotR})
are logarithmically UV-divergent.}.
The divergence is generated by the diagrams of Fig.~\ref{logmrhotR}; they give us 
the evolution of $c_T$  from $M_\rho$ to $\mc$.
Again, choosing the scale $M_\rho$ such that   $c_T(M_\rho)=0$, we have
\begin{equation}
\label{paramt3}
\xi c_{T}(\mc) = -\widehat{T}_{top}^{SM}2 c^{\prime\, 2}_{R} \xi^{2} \frac{1}{\epsilon_{t}}\log{(M_{\rho}^{2}/\mc^{2})}\, .
\end{equation}
Let us now integrate the custodians.
We are led  to the lagrangian Eqs.~(\ref{effcompoR}) and (\ref{effeleL})
with the coefficients
\begin{equation}
\label{relctLR3}
c_{R} = 0\ , \quad \tilde c_{L}^{(1)} =-\tilde c_{L}^{(3)}=\frac{1}{2}\, .
\end{equation}
As in the previous case, we have $c_R=0$  due to the custodial symmetry \cite{Agashe:2006at}. 
For $c_T$, the matching  at $\mu=\mc$ is given by Eq.~(\ref{match3})
where
\begin{equation}
\label{ttopr1}
\left[ \widehat{T}_{custodians} + \widehat{T}_{mix} \right]_{\mu \rightarrow \mc^{+}} = \widehat{T}_{top}^{SM} \left(  c^{\prime\, 2}_{R} \frac{\xi^{2}}{\epsilon_{t}} + 4c^{\prime\, 2}_{R}\xi^{2}
- 8 c^\prime_{R}\xi - \frac{16}{3} \epsilon_{t} \right).
\end{equation}
The running from $\mc$  to  $m_{t}$ proceeds by the same diagrams as those 
in Fig.~\ref{log} but with the replacements
$\tilde c_{R}\rightarrow c_R\xi/\epsilon_t$ and $c_{L}^{(1),(3)}\rightarrow \tilde c_{L}^{(1),(3)}\epsilon_t/\xi$. We obtain
\begin{equation}
\label{ttopru}
\xi c_{T}(m_{t}) = \xi c_{T}(\mc) + \widehat{T}_{top}^{SM} \left( -2\epsilon_{t} \right) \log \epsilon_{t}.
\end{equation}
Finally, when we match at the top mass scale, Eq.~(\ref{match2}), we get
\begin{equation}
\label{ttopr}
\left[ \widehat{T}_{top} \right]_{\mu \rightarrow m_{t}^{+}} = \widehat{T}_{top}^{SM} (-\epsilon_{t})\, .
\end{equation}
Again, we are not including the SM top contribution.
Adding Eqs.~(\ref{paramt3}), (\ref{ttopr1}), (\ref{ttopru}) and (\ref{ttopr}), we obtain  the total contribution to $\widehat T$  
\begin{align}
\label{Ttotal4}
\widehat{T} &= \widehat{T}_{top}^{SM} \left[ c^{\prime\, 2}_{R}\xi^{2} \left( \frac{1}{\epsilon_{t}} + 4 - \frac{2}{\epsilon_{t}} \log \frac{M_{\rho}^{2}}{\mc^{2}} \right) -8c^\prime_{R}\xi - \epsilon_{t} \left( \frac{19}{3} + 2 \log \epsilon_{t} \right) \right].
\end{align}
A plot of the value of $\widehat{T}$ is presented in Fig.~\ref{T1331} in the $\mc-c^\prime_{R}\xi/\xi_{R}$ plane for  $M_{\rho}=2.3$ TeV and $3.6$ TeV. 
We note  that the parameter $c^\prime_{R}$ is not related to any coefficient of the low-energy top lagrangian.  Nevertheless,
since one expects  $c^\prime_{R}\xi/\xi_R$ to be of order $1$
for a     composite $t_R$,    the  bounds from Fig.~\ref{T1331} 
can be considered indirect  limits  on the  degree of compositeness of  $t_R$.
These bounds  are  strong in the $c^\prime_{R}>0$ region, but  quite
weak for $c^\prime_{R}<0$. It is interesting to see that in
this latter region  it is  very natural to have a positive  contribution to $\widehat T$,
as needed for EWPT.

From the above analysis we  can  summarize  the following.
A composite $q_L$
is only likely in  case (a). 
It   yields to $c_L^{(1)}=-c_L^{(3)}\sim 1$ so it can be tested in modifications
of the top couplings.  
On the other hand,  a  composite $t_R$ is weakly constrained  in both cases.
Case (a) predicts a small  $\widehat T$, while in case (b) $\widehat T$
can receive sizable positive contributions, and therefore is favored 
by  EWPT.
Both cases, however, predict $c_R=0$,    so the only way to test this possibility 
is by  effects  coming from  $c_{4t}$ (four-top physics).

\subsection{One-loop contributions to $Zb\bar b$}

Although the coupling $Zb\bar b$ is not modified at the tree-level, it can
receive corrections at the one-loop level due to loops of SM particles and custodians that
break the custodial and $P_{LR}$ symmetry protecting  this coupling.
Here we only present the 
one-loop corrections to $Zb_L\bar b_L$   proportional to $c_{4q}$; they  are, as we will see,  the only one  that can  be  parametrically larger than the corrections  to  $\widehat T$,
and then can  put, in certain cases, stronger constrains on composite tops \footnote{This can be seen by inspection of the  one-loop diagrams contributing to $Zb_L\bar b_L$ in the  effective theory given in Section~\ref{effeclagra}. Loop diagrams  involving 
$c_{4q}$ and $c_L$  are quadratically divergent.}. 
In the limit $M_\rho\gg \mc\gg m_t$, we find, for the both cases of Eq.~(\ref{cases}),
\begin{equation}
\delta g_{b_{L}} = 
- \delta g_{b_{L}}^{SM} 3 c_{4q}\xi 
\left[c_{L}\xi \left( \frac{4}{\epsilon_t} \log{\frac{M_{\rho}^2}{\mc^2}} 
+ 4 \log{\epsilon_t} \right) 
+ 2 \log{\epsilon_t} \right]\, ,
\label{deltaz}
\end{equation}
where 
\beq
\delta g_{b_{L}}^{SM} = \frac{g}{\cos \theta_W} \frac{m_t^2}{16 \pi^2 v^2} \simeq 2 \cdot 10^{-3}\, , \nonumber
\eeq
corresponds to the top one-loop leading-contribution to $Zb\bar b$ in the SM.
Notice that Eq.~(\ref{deltaz}) shows contributions that grow with the custodian mass and  are logarithmically sensitive to the heavy resonance mass $M_\rho$.
Therefore, for a composite $q_L$, where $c_{4q}\sim c_L\sim 1$,
these contributions to 
$Zb\bar b$  can be  larger 
than those to $\widehat T$ for the case (a).
For example, for $c_{4q}\sim -1/6$, $c_L\sim -0.2$, $\xi\sim 1/4$, and $M_\rho\sim 2.3$ TeV, 
$\mc\sim 800$ GeV, the contributions to $\widehat T$ are below the experimental bound
but we find $\delta g_{b_{L}}/g_{b_{L}}\sim 0.013$ 
that is larger than the experimental constraint 
$-0.002 \lesssim\delta g_{b_{L}}/g_{b_{L}}\lesssim 0.006$.
These sizable contributions to $Zb\bar b$, however, scale with
$c_{4q}c_L\propto (y_L/g_\rho)^6$, while those to 
$\widehat T$  are proportional to $c_L^2\propto (y_L/g_\rho)^4$;
therefore the contributions to $Zb\bar b$ can be  parametrically suppressed with respect
to those to $\widehat T$ if $y_L$ is slightly smaller than $g_\rho$. 
For a composite $t_R$, contributions to $Zb\bar b$  proportional to 
the custodian mass or logarithmically sensitive to $M_\rho$ are not present, and therefore Fig.~\ref{T1331} will not suffer large modifications.

For very light custodians, the constraints 
from   $Zb\bar b$ can be as important  as those  from $\widehat T$
\cite{Carena:2007ua,Barbieri:2007bh}.
This implies that  the  allowed low-$\mc$  regions of Figs.~\ref{T11} and \ref{T1331} 
could be sligthly reduced 
by  the $Zb\bar b$ constraints.  We leave this calculation for a future publication.

\section{Phenomenological implications at future colliders}

In this section  we want to study  the experimental implications of having one of the top chiralities  being a composite state. 
For this  purpose, the effective lagrangian 
of section~\ref{effeclagra}
gives a useful  
model-independent parametrization of  the composite-top new  interactions.
We will not consider physics involving the Higgs that has been already studied in Ref.~\cite{Giudice:2007fh},
and we will only concentrate on top physics.

\subsection{Anomalous couplings}

The coefficients $c_{L}^{(1),(3)}$ and $c_R$ give rise to new contributions to the 
top coupling to the SM gauge bosons. In particular, for
the  $Zt_{L}\bar{t}_{L}$,  $Wt_{L}\bar b_{L}$  and $Zt_{R}\bar{t}_{R}$ couplings, we have
respectively
\begin{equation}
\frac{\delta g_{Zt_{L}t_{L}}}{g_{Zt_{L}t_{L}}} = \frac{(c_L^{(3)}-c_L^{(1)})\xi}{1-\frac{4}{3}\sin^{2}{\theta_W}}\ ,\quad\quad
\frac{\delta g_{Wt_{L}b_{L}}}{g_{Wt_{L}b_{L}}} = c_{L}^{(3)} \xi\ , \quad\quad
\frac{\delta g_{Zt_{R}t_{R}}}{g_{Zt_{R}t_{R}}} = \frac{3c_{R} \xi}{4 \sin^{2}{\theta_{W}}} \, .
\end{equation}
In the framework considered here we have 
$c_{L}^{(3)}\simeq -c_{L}^{(1)}$ and $c_R\simeq 0$, and therefore only
deviations on the $t_L$ couplings can be sizable.
To observe these deviations is not going to be easy.
At the LHC, top quarks are mostly produced in pairs  via the strong gluon fusion process $gg \rightarrow t\bar{t}$,  decaying  to $Wb$.    
To measure the  $Wt_Lb_L$ coupling,  however, a single top  must be mostly detected from the process 
$ ub \rightarrow d t$.  At the LHC this coupling  could  be measured  with a sensitivity around 7\% \cite{Weiglein:2004hn}, implying that   one could  see deviations if $c_L\xi\gtrsim 0.07$. 
For the $Zt\bar t$ coupling the situation is more difficult, since it  will not be able to be measured at the LHC. 
The   ILC, however,   will be the suitable   machine to
unravel the  composite nature of the top.
Studies show that the top couplings could be measured 
with an accuracy  as low as 1\% \cite{AguilarSaavedra:2001rg}.

\subsubsection{Subleading anomalous coupligs}

The operators of section~\ref{effeclagra} are the dominant ones in a $p^2/f^2$ expansion.
Nevertheless, there are other operators  that, although subleading, 
can have an important impact 
in future experiments. For a composite $t_R$
one of these subleading operators is
\begin{equation}
\label{WtRbR}
\frac{ic_{RR}}{f^2}\frac{y_b}{y_t}H^\dagger D_\mu \widetilde H \bar b_R\gamma^\mu t_R\, .
\end{equation}
where, due to the presence of the $b_R$, the coefficient of the operator is suppressed by the Yukawa coupling of the bottom $y_b/y_t\simeq 0.02$.
The coupling $c_{RR}$ is constrained by $b\rightarrow s\gamma$  
to be $c_{RR}\xi \lesssim 0.2$ \cite{Larios:1999au}. 
At the LHC this coupling  will be able to be tested  in top decays. 
Ref.~\cite{AguilarSaavedra:2006fy} gives 
a precision  $-3.2  \lesssim c_{RR}\xi \lesssim 6.8$ for an integrated luminosity 
of $L=10$ fb$^{-1}$.

Another subleading operator is 
\begin{equation}
\label{WW}
\frac{c_{M}y_t}{16\pi^2f^2}\bar q_LW^{\mu\nu} \widetilde H \sigma^{\mu\nu} t_R   \, ,
\end{equation}
where $W^{\mu\nu}$ is the field-strenght of the SM  $W$ boson.
Ref.~\cite{AguilarSaavedra:2006fy} gives a precision for this coupling at the LHC  of order $-3.6   \lesssim c_{M}\xi \lesssim 3.6$ for $L=10$ fb$^{-1}$.
Similar coupling for the 
gluon could be measured at  the LHC  with  an accuracy
$c_{M}\xi \simeq 0.4$ for $L=100$ fb$^{-1}$
\cite{Weiglein:2004hn}.

\subsection{Four-top interactions and  $pp \rightarrow t\bar{t}t\bar{t}(b\bar b)$}

\begin{figure}[t!]
\begin{center}
	\includegraphics[width=3in]{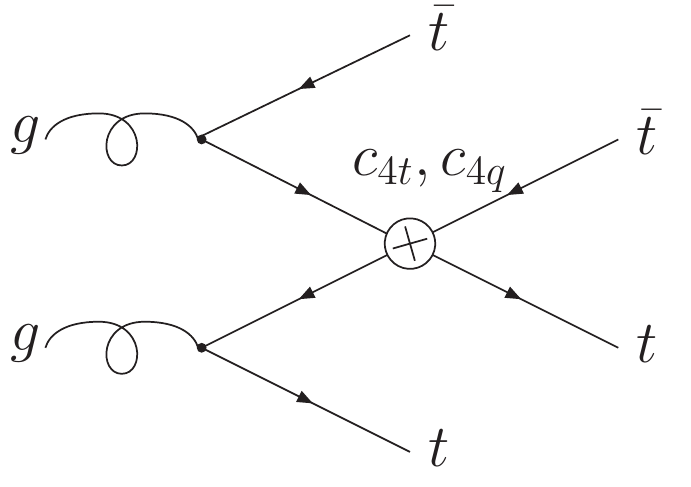}
\caption{Contribution of the four-top interaction  
to  the process $pp \rightarrow t\bar{t}t\bar{t}$.}
\label{4top}
\end{center}
\end{figure}

\begin{figure}[t!]
	\begin{center}
	\includegraphics[width=4in]{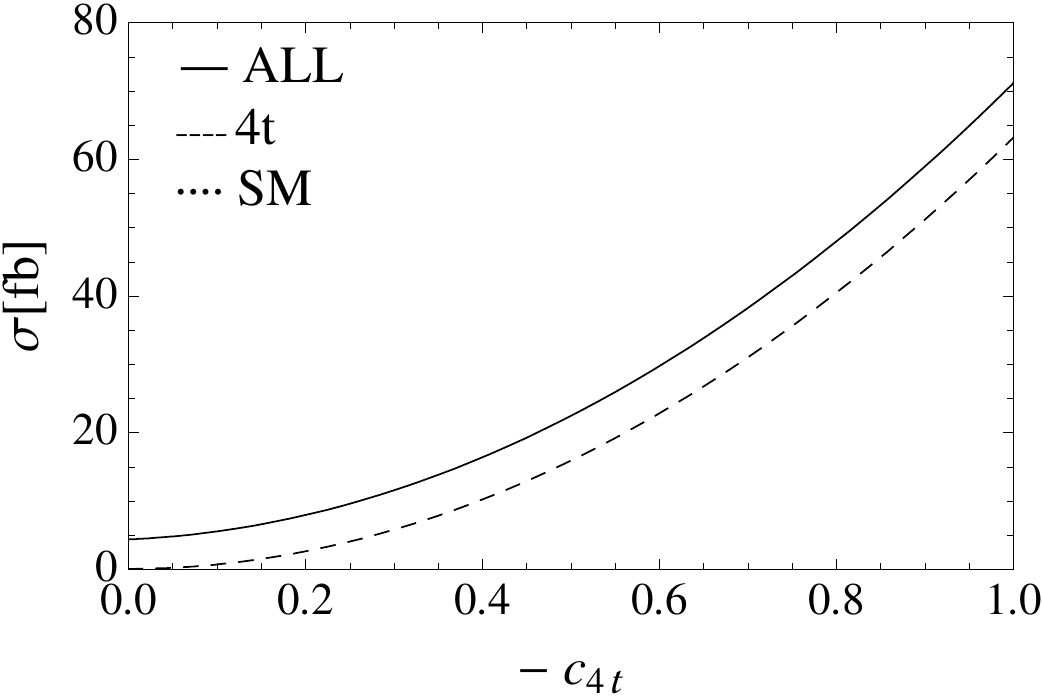}
	\caption{Cross-section for $pp \rightarrow t\bar{t}t\bar{t}$ as a function of $c_{4t}$ 	arising from  the operator ${\cal O}_{4t}$ 	(4t),   SM  diagrams  (SM) and  both
	(ALL).}
	\label{cdep}
	\end{center}
\end{figure}

The most genuine  effect of a composite top comes from 
the four-top interaction of Eqs.~(\ref{effcompoL}) and (\ref{effcompoR}). 
For a composite $t_R$ the operator ${\cal O}_{4t}=(\bar{t}_{R}\gamma^{\mu}t_{R})(\bar{t}_{R}\gamma_{\mu}t_{R})$  induces a  top-scattering amplitude that grows with the energy:
\begin{equation}
\label{tttt}
|{\cal A}(t_{R}\bar{t}_{R}\rightarrow t_{R} \bar{t}_{R})|^2=
64\frac{c^2_{4t}}{f^{4}}(u-2m_{t}^{2})^2\, .
\end{equation}
Similar expression  holds for a composite  $t_L$, induced in this case by  
the operator ${\cal O}_{4q}=(\bar{q}_{L}\gamma^{\mu}q_{L})(\bar{q}_{L}\gamma_{\mu}q_{L})$.
The growth with the energy of the four-top interaction
will lead  at the LHC  to an enhancement of the cross-section 
for $pp\rightarrow t\bar tt\bar t$  as shown in Fig.~\ref{4top}.
We have calculated 
the total cross-section for the process 
$pp\rightarrow t\bar tt\bar t$
using the MadGraph/MadEvent generator \cite{Maltoni:2002qb}.
For the computation we have used the CTEQ6M parton distribution functions and $Q = 1$ TeV as a reference value of the QCD renormalization and factorization scales. 
The result as a function of $c_{4t}$ is shown 
in Fig.~\ref{cdep}  for $f=500$ GeV. When 
the operator   ${\cal O}_{4t}$ is generated by a heavy color resonance,  
Eq.~(\ref{c4tq}), the   total cross-section for $pp\rightarrow t\bar t t \bar t$
is smaller than the SM one. 
Nevertheless, this  cross-section can be substantially  larger for larger values of $c_{4t}$.
Similar results have been presented previously in   Ref.~\cite{Lillie:2007hd}.

Due to  Eq.~(\ref{tttt}), we expect the $t\bar{t}$ pair coming from the four-top interaction  to  have a larger invariant mass and  transverse momenta than those coming from  gluons.
Hence, by taking $p_{T}(t_{1})>p_{T}(t_{2})$ (and the same for the anti-tops), we can identify 
the top $t_1$    as the scattered top and    the top $t_2$  as the spectator top.
We also expect the  $t_1\bar t_1$  pair to have large invariant mass $m$ and to be 
produced at large angles and then to have a small  pseudorapidity $\eta$.
These observables  can be useful  to discriminate the four-top signal versus  backgrounds.
\begin{figure}[t!]
	\begin{center}
	\includegraphics[width=3.3in]{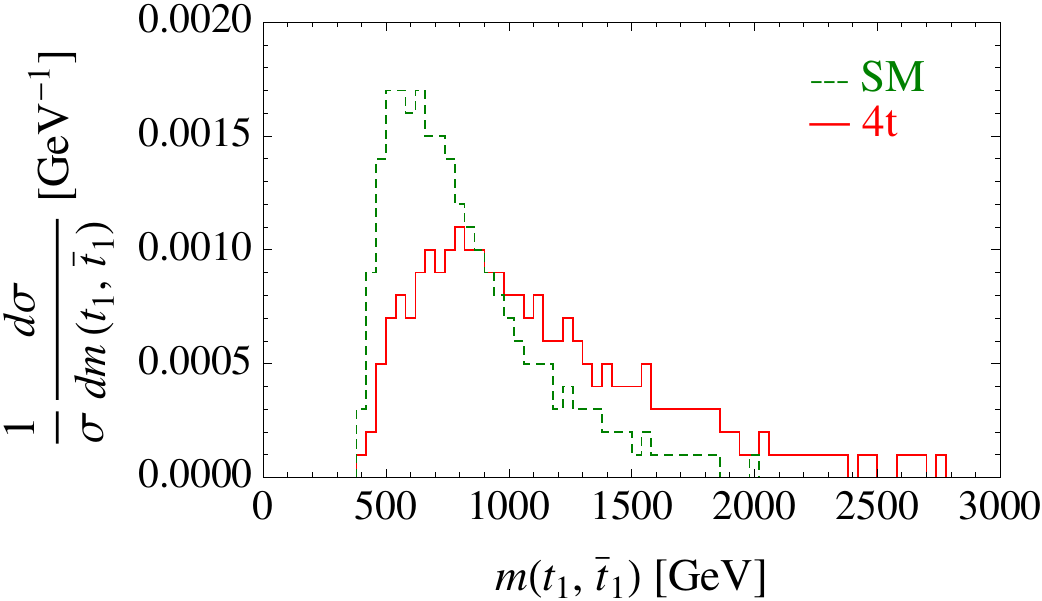}
	\includegraphics[width=3.3in]{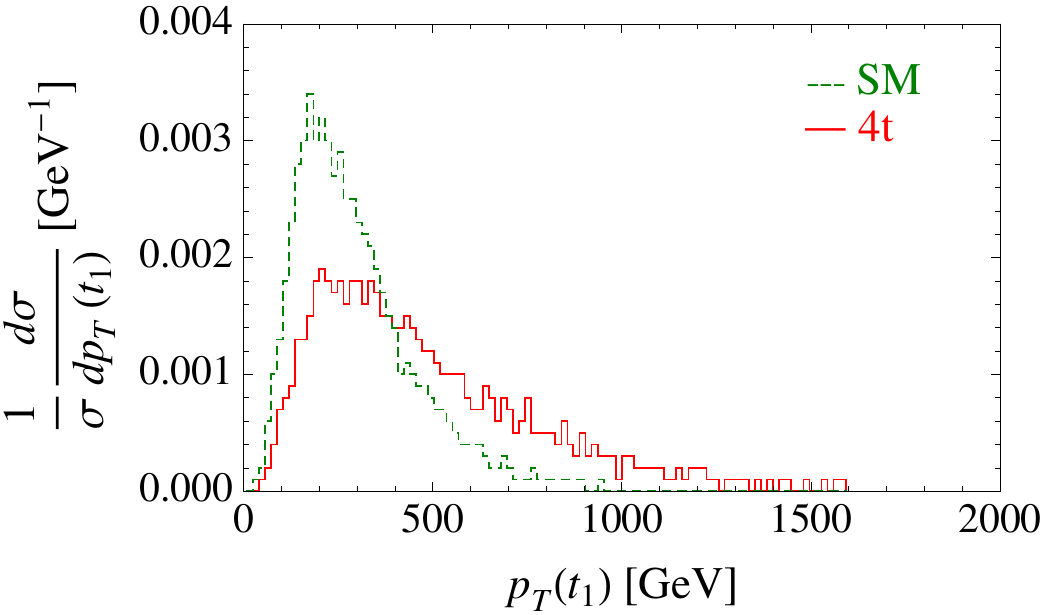}
	\includegraphics[width=3.3in]{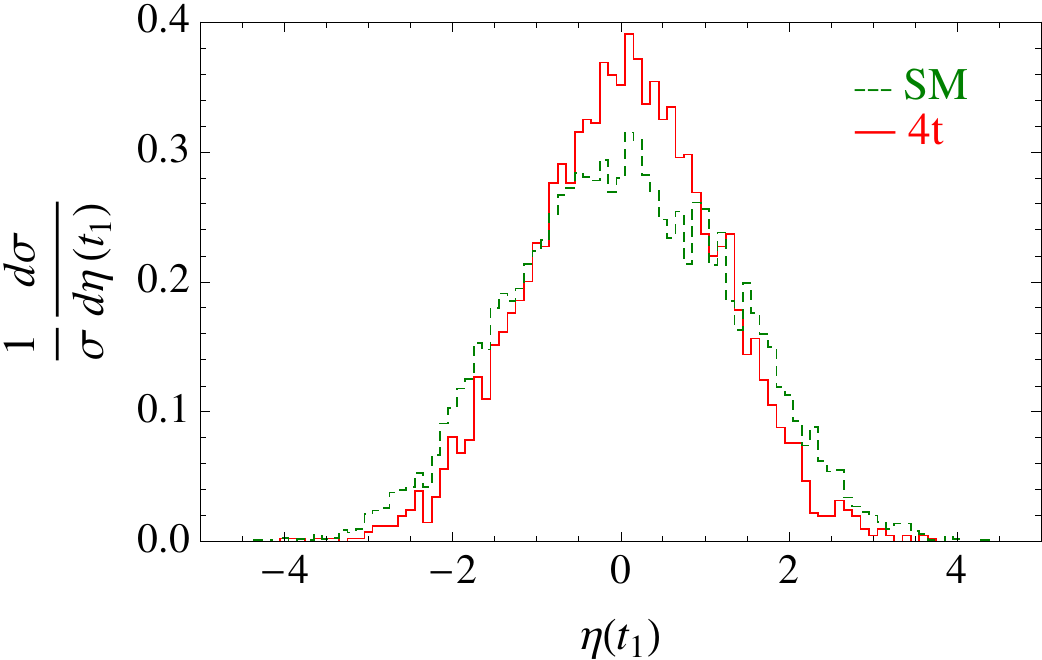}
	\caption{Normalized differential cross-section for $pp\rightarrow  t\bar tt\bar t$
	arising from the operator ${\cal O}_{4t}$  and the SM  plotted 
	 versus the invariant mass of the scattered top pair $m(t_{1},\bar{t}_{1})$,  the transverse momentum  $p_{T}(t_{1})$,  and  the  pseudorapidity $\eta(t_1)$.
	}
	\label{NPvsSMgraphics}
	\end{center}
\end{figure}

In Fig.~\ref{NPvsSMgraphics}
we plot the four-top  normalized differential cross-section  arising from the
four-top contact interaction,
and compare this with that of the SM.
We show   the
normalized differential cross-section versus the invariant mass 
of the scattered top pair $m(t_{1},\bar{t}_{1})$,  the transverse momentum of $t_1$,
$p_T(t_1)$,
and  its pseudorapidity $\eta(t_1)$; being normalized distributions,  they do not depend on  $c_{4t}$ or $f$. 
As expected, the  normalized differential cross-sections due to the new four-top 
contact interaction  are larger for large   $m(t_{1},\bar{t}_{1})$,  $p_T(t_1)$ or  small $\eta(t_1)$ than those of the SM.
In  Table~\ref{table1} we give the values of  the cross-section
for the four-top production 
for different cuts
in the  top-pair invariant-mass,   transverse momenta or  pseudorapidities.
We have taken  
$c_{4t}=-1/6$ and $f=500$ GeV, corresponding to the values of the  composite Higgs model,  Eqs.~(\ref{c4tq}) and (\ref{xiref}) respectively.
For the different cuts we give the value of the significance taken as
$
\mathcal{S}=\frac{\sigma_{ALL}-\sigma_{SM}}{\sqrt{\sigma_{SM}}}\sqrt{L},
$
where $L$ is the integrated luminosity that we take to be $L=100\textnormal{ fb}^{-1}$. 
We see that the  cuts do not substantially  increase  the significance. Nevertheless,
these cuts   can be useful in order to  eliminate reducible backgrounds, since 
the  detection of  the four tops   will  crucially depend  
on how well  one will be able to   reconstruct them at LHC.
Since   the scattered tops are  very energetic, 
their decay products will be highly collimated, making  
conventional reconstruction algorithms  difficult to apply.
In Ref.~\cite{Lillie:2007hd} an analysis at the particle level of the process $pp \rightarrow t\bar{t}t\bar{t}$ has been made, adopting the simple signature of at least two like-sign leptons $l^{\pm}l'^{\pm}$ plus at least two hard jets. 
They get  significances  $\sim 5$   for a value of $c_{4t}\sim 1/6$ 
and $f \sim 300-450$ GeV.
A more extended analysis at the detector level will be needed to 
study the feasibility of detecting this process.

\begin{table}[t!]
	\begin{center}
	\footnotesize{


\begin{tabular}{llccccccc}
\hline

\vspace{0.05in}
 & \textbf{Cuts} & $\mathbf{\sigma_{4t}}$\textbf{[fb]} &  $\mathbf{\sigma_{SM}}$\textbf{[fb]} & $\mathbf{\sigma_{ALL}}$\textbf{[fb]} & $\mathbf{\mathcal{S}}$ \\
 \vspace{0.05in}
 a) & no cuts & 1.8 & 4.6 & 7.0 & 11 \\
 \vspace{0.05in}
 b) & $m(t_{1},\bar{t}_{1})>650$ GeV & 1.5 & 2.8 & 4.5 & 10 \\
 \vspace{0.05in}
 c) & $p_{T}(t_{1}),p_{T}(\bar{t}_{1})>200$ GeV, $p_{T}(t_{2}),p_{T}(\bar{t}_{2})>30$ GeV & 1.3 & 2.2 & 3.5 & 8.7 \\
 \vspace{0.05in}
 d) & $|\eta|(t_{1}),|\eta|(\bar{t}_{1})<2$, $|\eta|(t_{2}),|\eta|(\bar{t}_{2})<4$ & 1.5 & 3.5 & 5.4 & 10 \\
 \vspace{0.05in}
 e) & (b) + (c) + (d) & 1.1 & 1.7 & 2.8 & 8.4 \\
 \hline
 \end{tabular} \\}
	\caption{Cross-section for $pp \rightarrow t\bar{t}t\bar{t}$ arising
	from ${\cal O}_{4t}$  with $c_{4t}=-1/6$ and $f=500$ GeV (4t),     SM  diagrams  (SM) and  both	(ALL) 	 for different cuts.
The corresponding significance ${\cal S}$ is also given.
	}
	\label{table1}
	\end{center}
\end{table}

\begin{table}[t!]
	\begin{center}
	\footnotesize{

\begin{tabular}{llccccccc}
\hline

\vspace{0.05in}
 & \textbf{Cuts} & $\mathbf{\sigma_{4q}}$\textbf{[fb]} & $\mathbf{\sigma_{SM}}$\textbf{[fb]} & $\mathbf{\sigma_{ALL}}$\textbf{[fb]} & $\mathbf{\mathcal{S}}$ \\
 \vspace{0.05in}
 a) & $p_{T}(b),p_{T}(\bar{b})>150$ GeV + $\Delta R (b,\bar{b})>1$ & 5.6 & 16 & 23 & 18 \\
 \vspace{0.05in}
 b) & (a) + $m(t,\bar{t})>600$ GeV & 3.9 & 6.0 & 11 & 19 \\
 \vspace{0.05in}
 c) & (a) + $m(b,\bar{b})>600$ GeV & 3.9 & 4.4 & 9.1 & 23 \\
 \vspace{0.05in}
 d) & (a) + $p_{T}(t),p_{T}(\bar{t})>300$ GeV & 1.3 & 1.2 & 2.6 & 13 \\
 \hline
 \end{tabular} \\}
	\caption{
	Cross-section for $pp \rightarrow t\bar{t}b\bar{b}$ 
	arising from ${\cal O}_{4q}$ with $c_{4q}=-1/6$ and $f=500$ GeV
		(4q),     SM  diagrams  (SM) and  both
	(ALL) 
		 for different cuts.
The corresponding significance ${\cal S}$ is also given.
	}
	\label{table5}
	\end{center}
\end{table}

\begin{figure}[t!]
	\begin{center}
	\includegraphics[width=3.3in]{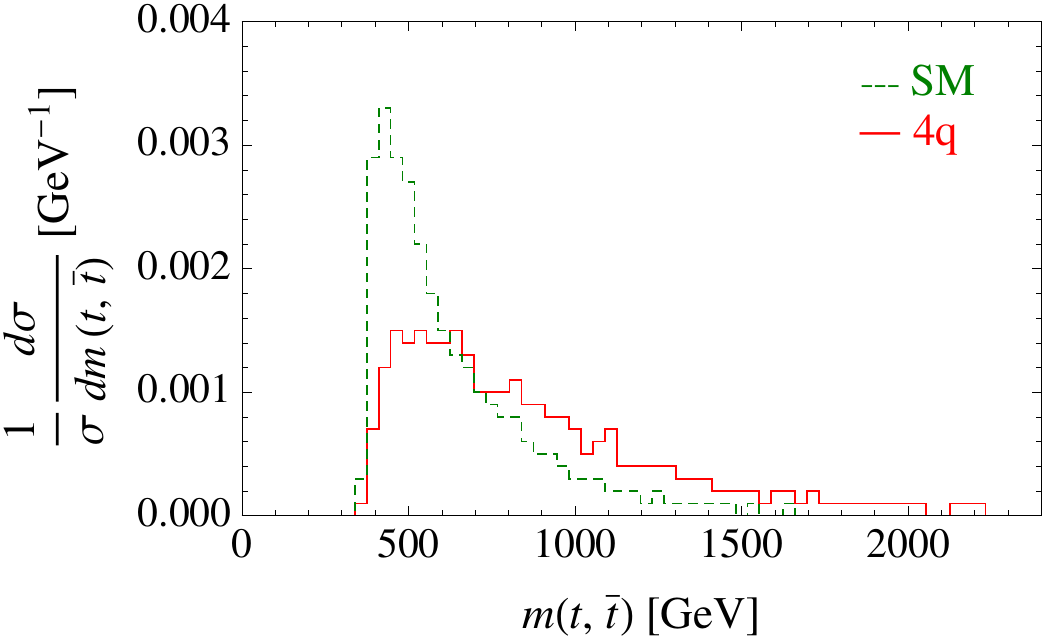}
	\includegraphics[width=3.3in]{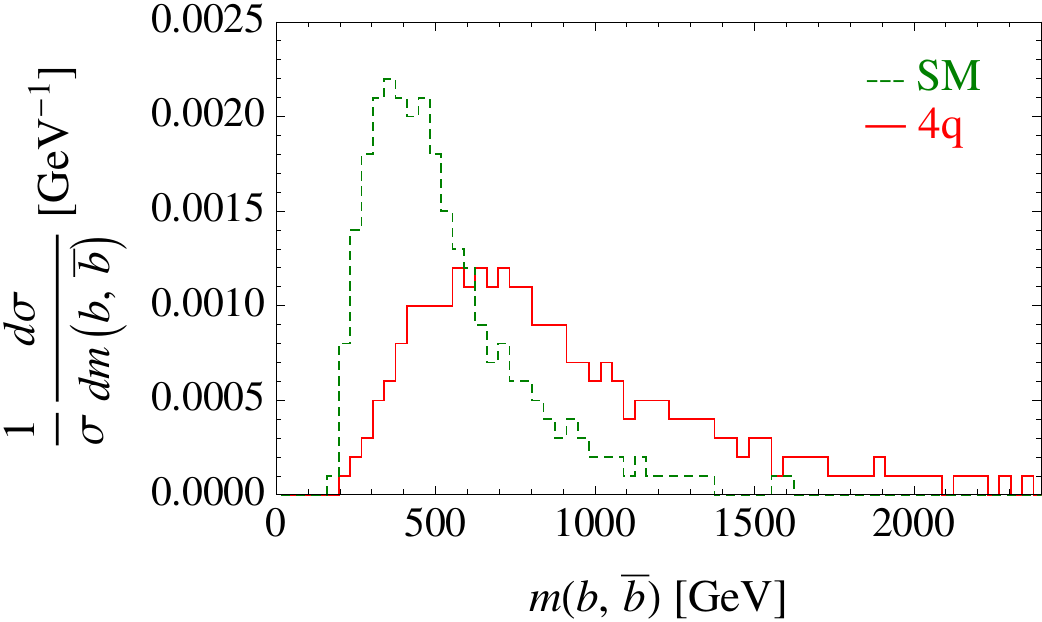}
	\includegraphics[width=3.3in]{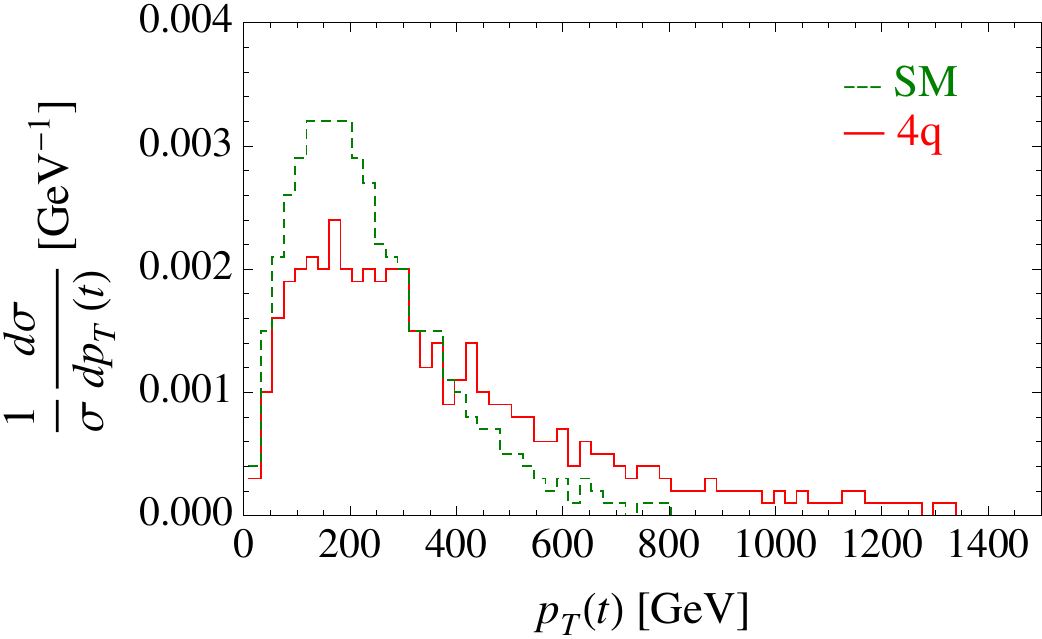}
	\includegraphics[width=3.3in]{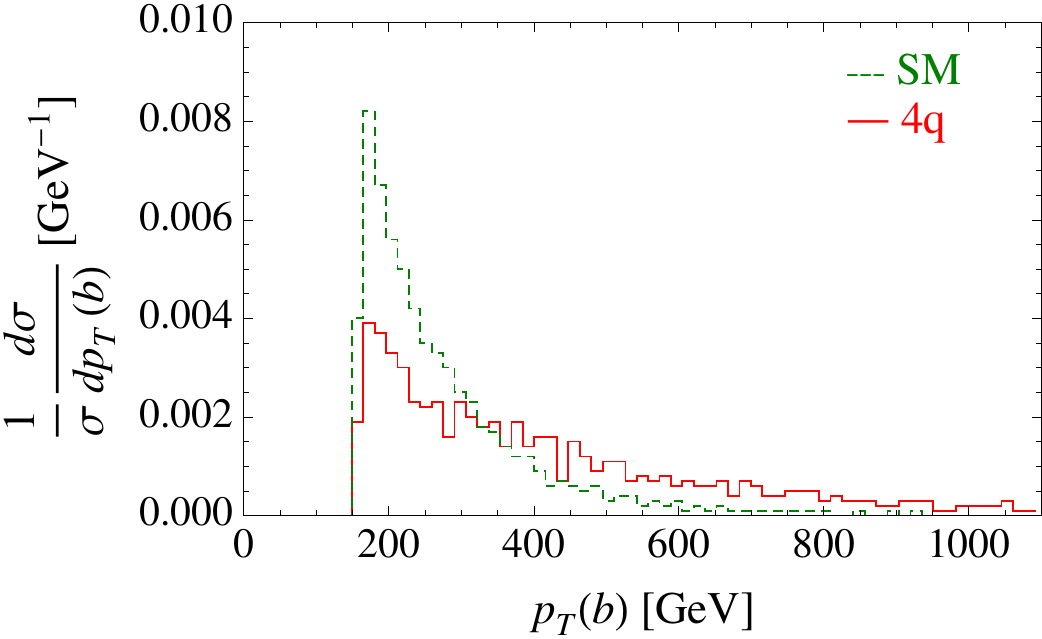}
	\includegraphics[width=3.3in]{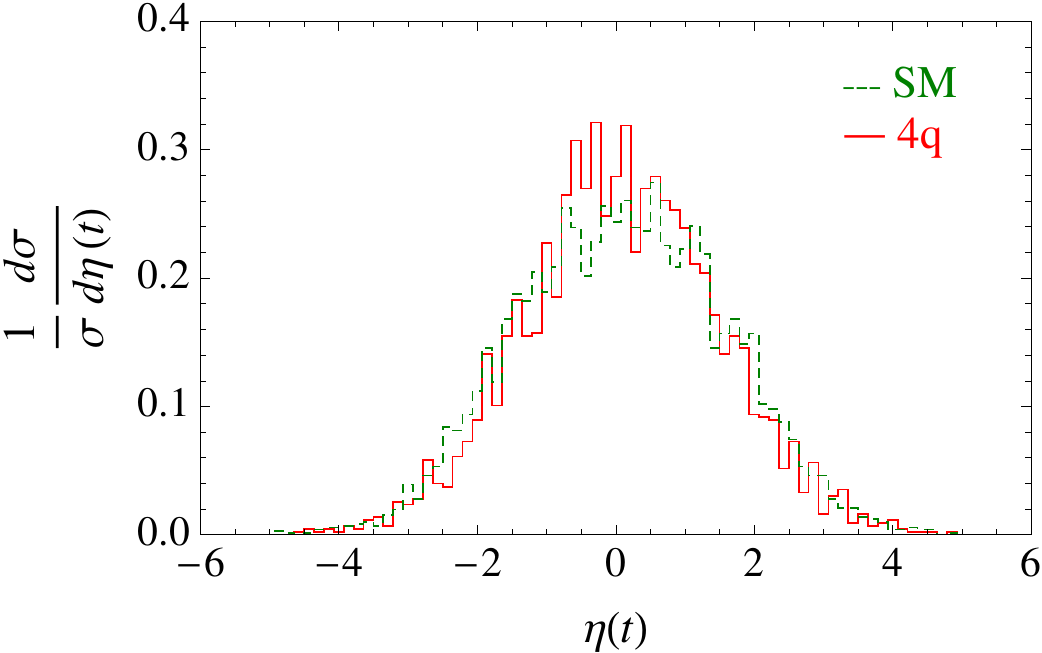}
	\includegraphics[width=3.3in]{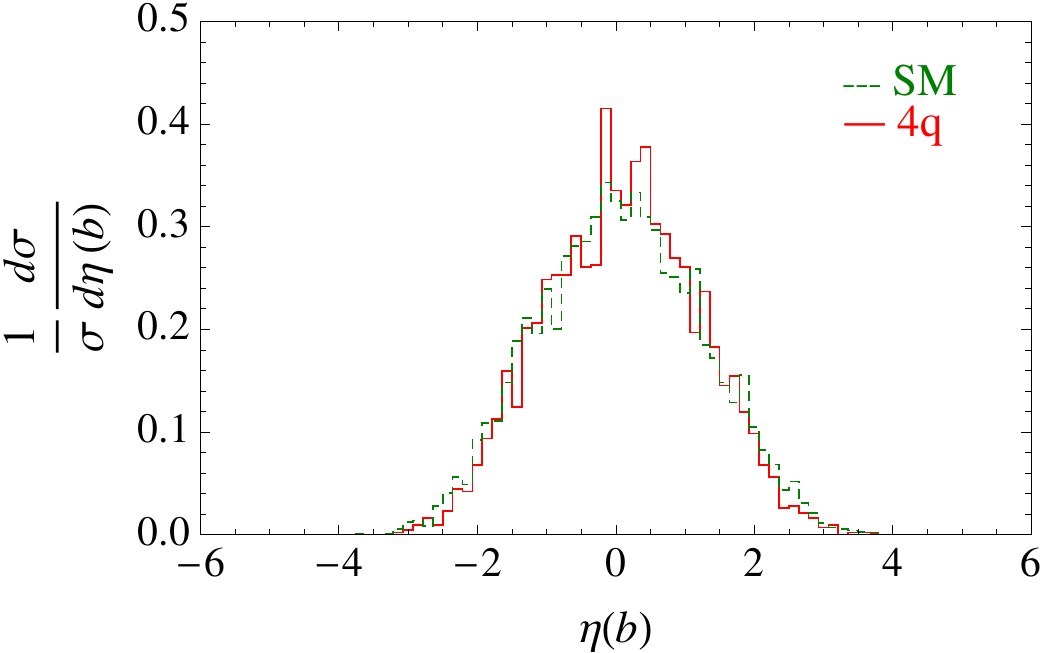}
	\caption{Normalized differential cross-section for
	$pp \rightarrow t\bar{t}b\bar{b}$ (with cuts $p_{T}(b), p_{T}(\bar{b}) > 150 \GeV$ and $\Delta R (b,\bar{b})>1$) induced by the operator
	${\cal O}_{4q}$ versus the invariant mass, the transverse momentum and the pseudorapidity of the tops or bottoms. We compare them with those of the SM.
	}
	\label{ttbb}
	\end{center}
\end{figure}

In the case of a composite $q_L$, the operator ${\cal O}_{4q}$
also induces  an  
amplitude for the process $b\bar b\rightarrow t\bar t$  
that grows with the energy:
\begin{equation}
\label{bbtt}
|{\cal A}(b_L\bar b_L\rightarrow t_L \bar t_L)|^2= 4\frac{c^2_{4q}}{f^{4}}(u-m_{t}^{2}-m_{b}^{2})^{2}\, .
\end{equation}
At the LHC this will give an enhancement of the cross-section 
of $pp\rightarrow t\bar t b\bar b$  similar to    Fig.~\ref{4top} but  with $b$ 
either as the spectator or the scattered quarks.
To  calculate with the MadGraph/MadEvent generator   the total cross-section
for $pp\rightarrow t\bar t b\bar b$ 
we will  demand a large $p_T$ for the bottom quarks and a large separation angle between them,
in order to avoid large logarithmic corrections due to collinear $b\bar b$ coming from the gluon \cite{Bernreuther:2008ju}
\footnote{Alternatively, we could  sum up these large logarithmic terms
by introducing the b-quark PDF and calculating the process  $pp\rightarrow t\bar t$. 
Nevertheless, the huge SM contribution to top-pair production would in this case swamp the effect of a composite top   coming from  Eq.~(\ref{bbtt}).
We thank Tim Tait for pointing out these problems to us.}.
In  Table~\ref{table5} we  give  
the cross-section  for 
$pp\rightarrow t\bar t b\bar b$  for 
$p_T(b),p_T(\bar b)>150$ GeV and $\Delta R(b,\bar b)=\sqrt{(\eta_b-\eta_{\bar b})^2+(\phi_b-\phi_{\bar b})^2}>1$ where $\phi_i$ is the azimuthal angle
(we take the renormalization scale  $Q=0.5$ TeV,  $c_{4q}=-1/6$ and $f=500$ GeV).
To show the dependence of the  $t\bar t b\bar b$ production cross-section
versus  
the invariant mass, transverse momentum and pseudorapidity of the bottom and top,
we plot in
Fig.~\ref{ttbb}  the  normalized differential cross-sections for 
$pp\rightarrow t\bar t b\bar b$ induced by
the four-fermion  interaction,
and  compare them with the SM ones.
The variation  of the cross-section and the significance of the signal for several cuts 
is given in Table~\ref{table5}.

\subsubsection{Top polarization measurement}

The determination of the top-quark polarization 
gives a complementary   way to probe the properties of the top interactions and to discriminate between either right-handed or left-handed top compositeness. 
At the LHC, the top quarks   are dominantly produced  unpolarized 
by QCD interactions. 
In the presence of the operators ${\cal O}_{4t,4q}$, however,
the $t\bar{t}t\bar{t}$ production
yields   an excess of either right- or left-handed scattered tops  that can be 
visible by  measuring  the top polarization.

\begin{figure}[t!]
	\begin{center}
	\includegraphics[width=4in]{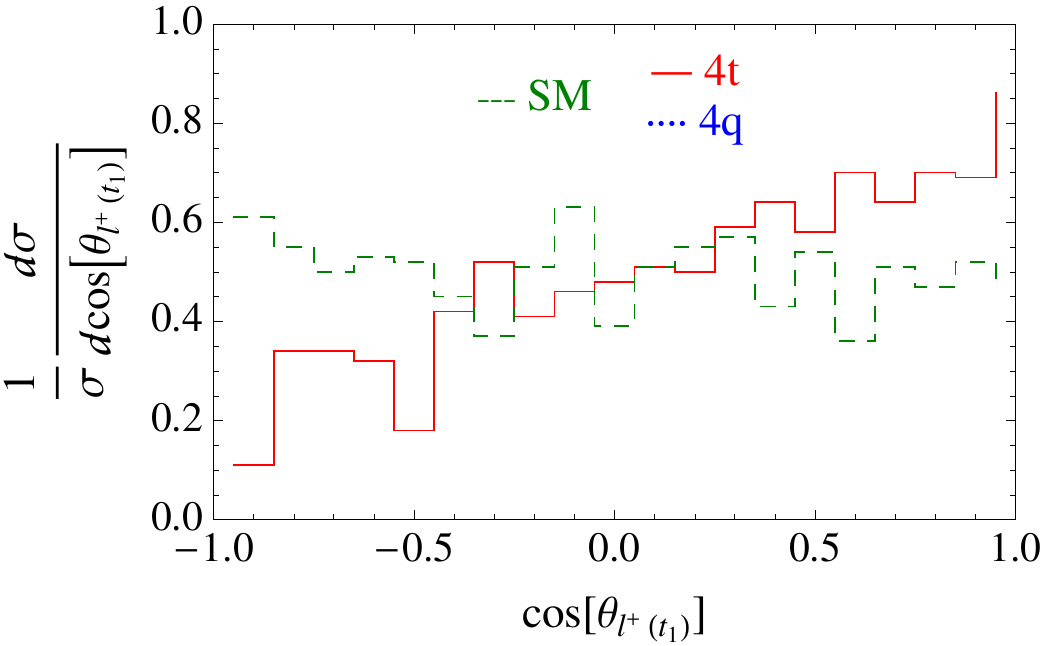}
	\caption{Normalized differential cross-section for $pp \rightarrow t\bar{t}t\bar{t}$  versus $\cos{\theta_{X}}$
	where $X$ is the  lepton coming from the decay of the scattered top.
	}
	\label{red}
	\end{center}
\end{figure}

The   polarization of the top quarks can be analyzed
from   the    angular distribution of their  decay products. 
In the decay channel $t \rightarrow W^{+}b \rightarrow l^{+}\nu b, q\bar{q}'b$, the angular distribution of the ``spin analyzers'' $X = l^{+}, \nu, q, \bar{q}', W^{+}, b$ 
is given by
\begin{equation}
\label{spin}
\frac{1}{\Gamma} \frac{d\Gamma}{d\cos{\theta_{X}}} = \frac{1}{2} (1+\alpha_{X} \cos{\theta_{X}})\, ,
\end{equation}
with $\theta_{X}$ being the angle between the direction  of $X$ (in the top rest frame)
and the direction of the top polarization.  The constants $\alpha_{X}\in[-1,1]$, 
take in the SM  the approximate values   $\alpha_{l^{+}} = \alpha_{\bar{d}} = 1$, $\alpha_{\nu} = \alpha_{u} = -0.32$, 
 $\alpha_{W^{+}} = - \alpha_{b} = 0.41$ \cite{Hubaut:2005er}. 
 From Eq.~(\ref{spin}) we can  obtain the top production 
 differential cross-section
\begin{equation}
\label{topspin}
\frac{1}{\sigma}\frac{d\sigma}{d\cos\theta_X}=F_{R}+F_{L}=\frac{A}{2} (1+\alpha_{X}\cos{\theta_{X}}) + \frac{1-A}{2} (1-\alpha_{X}\cos{\theta_{X}}),
\end{equation}
where $F_{R}$ and $F_{L}$ are respectively the angular distributions for right- and left-handed quarks and $A$ corresponds to   the fraction of right-handed quarks produced (therefore $A \in [0,1]$). In the SM we expect $A\sim 1/2$.
In Fig.~\ref{red} we show the normalized differential cross-section for four-top production at the LHC
as a function of
$\cos\theta_{X}$  where $X=l^+$ is  the lepton coming from the  top with the highest $p_T$.
We show this for  tops     arising   either  
from ${\cal O}_{4t}$ (4t)  or   ${\cal O}_{4q}$ (4q), and 
compare   with the SM case.
By fitting Fig.~\ref{red} with the   distribution  Eq.~(\ref{topspin})  
we find 
$A\simeq 0.5$   for the SM, while $A\simeq 0.8$ and $0.2$ respectively 
for the 4t and 4q case.
From Eq.~(\ref{topspin}) one can calculate 
forward-backward  asymmetries  in the  lepton channel similar to those of  
Ref.~\cite{Agashe:2006hk}
that can be useful 
to disentangle the helicity  of the top
if an excess in the  four-top production is found at the LHC.

\section{Conclusions}

In  models  in which   EWSB is  triggered by a new  strong sector or  a warped extra dimension,
the SM fermions can get their masses by mixing with  composite states (or operators)
of the  new sector.
In this framework it is  natural to consider due to the heaviness of the top
that one of its chiralities, $q_L$ or $t_R$,   is mostly composite.

In this article we have seen that present experimental bounds do not 
rule out  this possibility.
The custodial symmetry of the BSM sector plays an important role guaranteeing 
that the $T$-parameter  and $Zb\bar b$ do not get corrections at tree-level
for the cases (a) and (b) of Eq.~(\ref{cases}).
We have calculated the one-loop effects to the $T$-parameter
and showed, for a composite $q_{L}$,
that while in  the case (b) 
the bounds from  $\widehat T$
are very restrictive (Fig.~\ref{T13}), for the case (a), 
the presence of 
the custodial partners of the top, the  custodians, 
avoids large one-loop contributions  to  $\widehat T$    (Fig.~\ref{T11}).
For a composite $t_R$ the bounds from $\widehat T$ are 
very weak;   case (a) does not generate contributions to $\widehat T$, while  for  case (b)
one finds wide  allowed regions (Fig.~\ref{T1331}).
Our one-loop calculation   shows that  moderate and positive 
contributions to $\widehat T$ are more probable  in regions in which
the coefficients of the higher-dimensional operators $c_i$ are negative.
These regions, although absent  in minimal holographic models  \cite{Carena:2007ua}, can be present in more  generic scenarios.
These positive contributions to $\widehat T$ are needed in this class of models 
in order to accommodate a generic positive contribution to the 
$S$-parameter.

At future accelerators, we have seen that top compositeness  
can be tested  by looking for deviations on the $Zt\bar t$ and $Wt\bar b$ coupling.
Only the second one, however, can be measured with certain accuracy at the LHC.
The ILC  would clearly be an excellent  machine to probe the properties of the top
and determine its degree of compositeness. 
A second  important effect of top compositeness is 
the presence of four-top contact terms that  enhances the cross-section for
$pp\rightarrow t\bar t t\bar t$  at high-energies.
We have calculated the cross-section of this process  at the LHC
for  the case  of  a composite $t_R$, and showed several observables that can allow us
to discriminate   from  the  SM prediction.
It is however unclear, due to the smallness of the cross-section, whether the four-top production  can be seen at the LHC.  
Clearly, a  more detailed analysis is  needed to assure the feasibility   of this process.
Similar analysis has been discussed  for  the process  $pp\rightarrow t\bar t b\bar b$
for the case of a composite $q_L$.

We finalize saying that the composite nature of the top
could also be seen indirectly by detecting the custodians.
Studies in this direction have been recently carried out in 
Ref.~\cite{Contino:2008hi}.

\section*{Acknowledgments}

This work  was  partly supported   by the
FEDER  Research Project FPA2005-02211
and DURSI Research Project SGR2005-00916.
The work of J.S. was also  supported by 
the Spanish MEC FPU grant  AP2006-03102.


\end{document}